\documentclass{aa}

\usepackage{float}
\usepackage{threeparttable}
\usepackage{graphicx,rotating,epsfig,color}
\usepackage{times,latexsym,txfonts,natbib}
\usepackage{longtable,lscape,multirow,supertabular,subfigure,multirow}

\def\Mdot{\mbox{\.M}}

\newcommand  \HII{\ion{H}{ii}}
\newcommand  \m{\mathrm}

\bibpunct{(}{)}{;}{a}{}{,} 
\begin{document}
\title{A bimodal dust grain distribution in the IC 434  \HII\ region}
\subtitle{}
\author{B.B. Ochsendorf \& A.G.G.M. Tielens}
\institute{Leiden Observatory, Leiden University, P.O. Box 9513, NL-2300 RA, The Netherlands \\ 
\email{ochsendorf@strw.leidenuniv.nl}\label{inst1}}

\abstract
{Studies of dust evolution and processing in different phases of the interstellar medium is key to understanding the lifecycle of dust in space. Recent results have challenged the capabilities and validity of current dust models, indicating that the properties of interstellar dust evolve as it transits between different phases of the interstellar medium.}
{We characterize the dust content from the IC 434 \HII\ region, and present a scenario that results in the large-scale structure of the region seen to date.}
{We conduct a multi-wavelength study of the dust emission from the ionized gas, and combine this with modeling, from large scales that provide insight into the history of the IC 434/L1630 region, to small scales that allowe us to infer quantitative properties of the dust content inside the \HII\ region.}
{The dust enters the \HII\ region through momentum transfer with a champagne flow of ionized gas, set up by a chance encounter between the L1630 molecular cloud and the star cluster of $\sigma$ Ori. We observe two clearly separated dust populations inside the ionized gas, that show different observational properties, as well as contrasting optical properties. Population A is colder ($\sim$ 25 K) than predicted by widely-used dust models, its temperature is insensitive to an increase of the impinging radiation field, is momentum-coupled to the gas, and efficiently absorbs radiation pressure to form a dust wave at 1.0 pc ahead of $\sigma$ Ori AB. Population B is characterized by a constant [20/30] flux ratio throughout the \HII\ region, heats up to $\sim$ 75 K close to the star, and is less efficient in absorbing radiation pressure, forming a dust wave at 0.1 pc from the star.}
{The dust inside IC 434 is bimodal. The characteristics of population A are remarkable and can not be explained by current dust models. We argue that large porous grains or fluffy aggregates are potential candidates to explain much of the observational characteristics. Population B are grains that match the classical description of spherical, compact dust. The inferred optical properties are consistent with either very small grains, or large grains in thermal equilibrium with the radiation field. Our results confirm recent work that stress the importance of variations in the dust properties between different regions of the ISM.}

\titlerunning{}
\maketitle

\section{Introduction}

The advent of high-resolution infrared (IR) imaging during the last decade has provided indisputable evidence that dust resides within the ionized gas of \HII\ regions, rather than being associated with photo-dissociation regions along the line of sight \citep[e.g.,][]{povich_2007,paladini_2012}. More than often, the discussion in literature is focussed on the question of what kind of dust population is able to survive inside an \HII\ region, as the physical conditions in these regions predict that dust will be removed and/or destroyed in the ionized gas, either through radiation pressure, a stellar wind, dust sublimation, or dust sputtering \citep{inoue_2002, everett_2010, draine_2011}. 

The work of \citet{chini_1986} triggered the thought of a separate dust component inside \HII\ regions, which was supported by the model presented in \citet{everett_2010}, who discussed the possibility of evaporating cloudlets overrun by the ionization front (IF), resupplying the ionized gas with a new generation of dust. \citet{paladini_2012} showed that the IR emission from evolved \HII\ regions is typically stratified as follows: polycyclic aromatic hydrocarbons (PAHs; \citealt{tielens_2008}) and cold dust ($\sim$ 20 K) trace the PDR and a dense shell of the \HII\ region, respectively, while 24 $\mu$m emission peaks inside the ionized gas, close to the central source. \citet{ochsendorf_2014b} proposed a new scenario to simultaneously explain the presence and morphologies of the mid-IR emission inside \HII\ regions. Whereas several authors attribute the 24 $\mu$m emission inside ionized regions to the increase of VSGs compared with BGs \citep{flagey_2011,paradis_2011}, the analysis of the IR arc in IC 434 \citep{ochsendorf_2014a} (from now: Paper 1) revealed that the increased heating by stellar photons from the nearby star can explain the mid-IR emission. 

Besides the intrinsic nature of the dust inside \HII\ regions, recent studies (\citet{salgado_2012}, Paper 1) have indicated that dust properties inside \HII\ regions may differ from that used in widely-used dust models of the diffuse interstellar medium (ISM) \citep{li_2001,compiegne_2011}. In addition, results from the {\em Planck} mission indicate that dust properties evolve when transiting from one region of the ISM to another \citep[e.g.,][]{abergel_2013}. Here, we focus on dust grains entrained within photo-evaporation flows, and follow the evolution as they transit from the molecular cloud phase to ionized regions of space.

We examine the dust content and processes governing the properties of the dust inside \HII\ regions, using the IC 434 region as a benchmark model. The multitude of observations available for the region, its large angular extent, and its profitable location well outside the Galactic plane, makes IC 434 a suitable candidate to pursue this question. We extend the work and analysis described in Paper 1 to the far-IR. The new observations allow us to probe the dust content of the IC 434 region on a wide range of wavelengths which reveals a new dust population in the \HII\ region. When combined with the theoretical model described in Paper 1, the analysis provides a means to constrain the properties of the dust entrained in the ionized gas flow of the IC 434 region. In particular, we will argue that the dust content in IC 434 is best explained through a bimodal distribution, which differs significantly from that seen in the diffuse ISM. We discuss our findings in the light of dust evolution inside the molecular cloud from which the dust is evaporated, such as the formation of aggregates through coagulation, and through recent findings of dust evolution as constrained by, e.g., the {\em Planck} mission.

In Sec. \ref{sec:1bshell}, we set the stage by digging into the history of the large-scale structure of the IC 434/Orion B molecular cloud region, that provides us with an updated and improved model for the ionized gas flow in IC 434 (as compared to Paper 1), necessary to derive the properties of the dust introduced into the \HII\ region. In Sec. \ref{sec:observations}, we present the observations. After characterizing the global properties of the dust, we combine the observations with detailed modeling in Sec. \ref{sec:introducingdust}, to derive properties of the emitting grains coming from the ionized gas. The results of the modeling are described in Sec. \ref{sec:results}. We discuss our findings in Sec.\ref{sec:discussion}, and conclude in Sec. \ref{sec:conclusions}.

\section{Observations}

The observations used in this work are in part described in Paper 1, but are extended with {\em Herschel} observations at $\lambda$ $\textgreater$ 70 $\mu$m, {\em Planck} observations \citep{planck_collaboration1_2014}, and the {\em Planck} dust model R1.20 \citep{abergel_2013}. PACS 160 $\mu$m and SPIRE 250 $\mu$m, 350 $\mu$m and 500 $\mu$m photometry were taken from the Herschel Science Archive (HSA) from the Gould Belt survey \citep{andre_2010} with observation IDs 1342215984 and 1342215985. {\em Herschel} data was inter-calibrated with IRAS 70 $\mu$m (PACS 70 $\mu$m), {\em Planck} 857 GHz (PACS 160 $\mu$m, SPIRE 250 $\mu$m and 350 $\mu$m), and Planck 545 GHz (SPIRE 500 $\mu$m) through the method described in \citet{bernard_2010}. The {\em Herschel} resolutions are 8.4" (70 $\mu$m), 13.5" (160 $\mu$m), 18.2" (250 $\mu$m), 24.9" (350 $\mu$m), and 36.3" (500 $\mu$m), respectively, for parallel mode observations at (60" s$^{-1}$) scan rates. 

The reduction of Spitzer/IRS observations is described in Paper 1, but is extended here with the Short-Low (SL) modules, covering the 5 - 15 $\mu$m region. The IRS SL detector suffered from a background level, causing apparent gradients throughout the modules and mismatches between the SL1 and SL2 spectra. This background level persisted after background subtraction of the BCD images and affected the reduced spectra. Following \citet{sandstrom_2012}, we opted to correct the BCD images with the use of the inter-order regions that resulted in a significant improvement of the quality of the spectra.

\section{The expanding HI shell surrounding Orion OB1b}\label{sec:1bshell} 

In this section, we connect the IC 434 \HII\ region with the large-scale structure of the Orion Belt region. We will argue that the large IR bubble, approximately centered on the open cluster of $\sigma$ Ori, is blown by the Belt star $\epsilon$ Ori, and that $\sigma$ Ori has crossed the distance from the Orion OB1c region, currently approaching the L1630 molecular cloud. Besides that the analysis provides a new insight into the history and evolution of the Belt region, the argumentation ultimately leads to the development of the flow model described in Sec. \ref{sec:introducingdust} and, subsequently, the derivation of grain properties that are entrained within the ionized gas.

\begin{figure*}
\centering
\includegraphics[width=18cm]{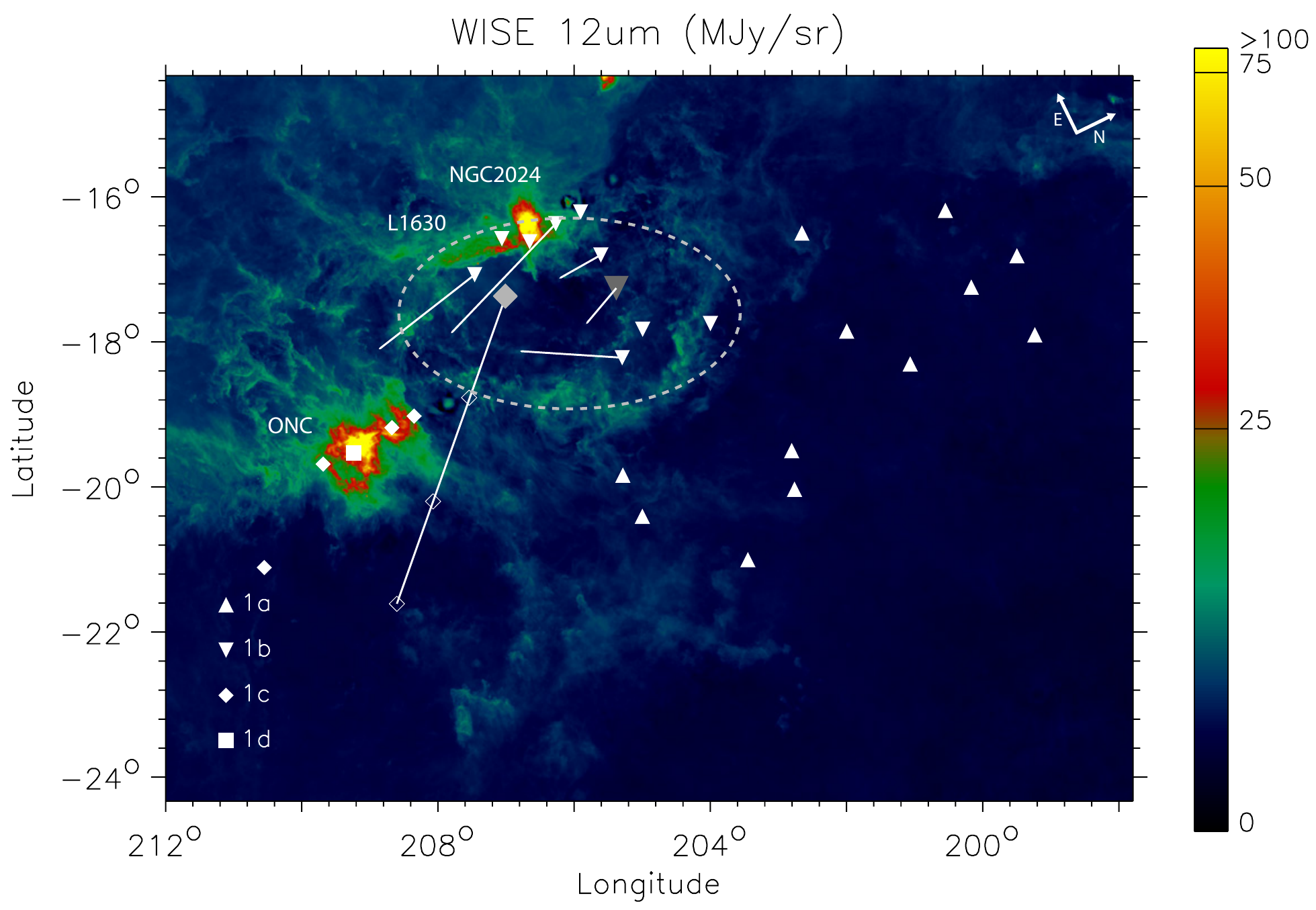} 
\caption{The WISE mid-IR view of the region surrounding the Orion OB association at 12 $\mu$m. The different sub-groups of the Orion OB association are overplotted for stars of spectral type B2 and earlier \citep{brown_1994}, where the oversized symbols represent $\sigma$ Ori (diamond) and $\epsilon$ Ori (inverted triangle), respectively. The 12 $\mu$m emission reveal the edges of the Orion molecular clouds as they are being illuminated by the ionizing flux of members of the Orion OB association. Also shown are several prominent structures in the image: the L1630 molecular cloud, the emission nebula NGC2024, and the Orion Nebular cloud (ONC). Of particular interest here is the bubble centered at ($l$,$b$) = (206.1,-17.5), which is the IR counterpart of the GS206-17+13 HI shell \citep{ehlerova_2005} and encircles the members of Orion OB1b (dashed outline). Overplotted in solid lines are the space motions of the stars, tracing {\em back} the motion of the stars in the plane of the sky during 1 Myr. For $\sigma$ Ori, this trajectory has been extended to 3 Myr, where the open diamond symbols mark 1 Myr, 2 Myr, and 3 Myr.} 
\label{fig:shell}
\end{figure*}

\subsection{The GS206-17+13 shell: mass and kinetic energy}

The HI shell GS206-17+13 \citep{ehlerova_2005}, centered at ($l$,$b$) = (206.1,-17.5), has a central radial velocity of $v_\m{LSR}$ = 12.8 km s$^{-1}$ and an expansion velocity of roughly $v_\m{exp}$ = 8 km s$^{-1}$. It extends $\sim$ 2$^{\circ}$ $\times$ 4$^{\circ}$ (13 $\times$ 26 pc at 385 pc; see below) in HI, and is also seen at mid- to far-IR wavelengths (Fig. \ref{fig:shell}). Apparently, it is centered on the Orion OB1b association and $\sigma$ Ori. The total mass in the shell is estimated by adopting a dust emissivity per unit mass of gas, $\kappa_\m{\nu}$, which is often modeled as a power law in the form of $\kappa_\m{\nu}$ = (10)$v$$^{\beta}$, with the frequency $\nu$ in THz \citep{beckwith_1990} and $\beta$ the spectral index of the dust opacity. It should be noted that this relation could be more complex, depending on the exact grain properties (e.g., \citet{abergel_2013}, and references therein). We use the $\tau_\m{353}$ (optical depth at 353 GHz) and the $\beta$ maps from the Planck dust model R1.20 \citep{abergel_2013}, and define an aperture that encloses the bubble shell seen in WISE and Planck images. We estimate the contribution of Galactic emission along the line-of-sight unrelated with the bubble shell by determining the value of $\tau_\m{353}$ in a region close by at the same Galactic latitude ($\sim$ 23\% of the total emission), and subtract this from the optical depth measured within the aperture. With $\kappa_\m{\nu}$ =  2.1 cm$^{-2}$ g$^{-1}$ at 353 GHz, and $\beta$ = 1.5 as constrained by the Planck model, we calculate the total dust mass of the shell $M_\m{d}$ = ($\tau_\m{353}$/$\kappa_\m{\nu}$)$S$, with $S$ the physical size of the emitting region. We find a total dust mass of $M_\m{d}$ = 34 $M_\m{\odot}$ or, equivalently, a gas mass of $M_\m{g}$ = 3400 $M_\m{\odot}$ (using a gas-to-dust ratio $R$ = 100). By distributing the mass back into a homogeneous sphere with radius 20 pc (the average radius of the HI shell), we find an average density of $n_\m{local}$ = 4.1 cm$^{-3}$, i.e., the original density in which the bubble expanded. Then, the total kinetic energy of the shell can be estimated with the expansion model described in \citet{chevalier_1974}:

\begin{equation}
\label{eq:g0}
\frac{E_\m{tot}}{\m{erg}} = 5.3 \times 10^{43} \left( \frac{n_\m{local}}{\m{cm^{-3}}} \right)^{1.12} \left( \frac{r_\m{sh}}{\m{pc}} \right)^{3.12} \left(\frac{v_\m{exp}}{\m{km\ s^{-1}}} \right)^{1.4}
\end{equation} 

\noindent Here, $r_\m{sh}$ is the observed radius of the shell. Using $n_\m{local}$ = 4.1 cm$^{-3}$, $r_\m{sh}$ = 20 pc and $v_\m{exp}$ = 8.0 km s$^{-1}$, we find $E_\m{tot}$ = 5 $\times$ 10$^{49}$ erg, which is only a factor of few lower than the typical kinetic energy injected by a SN ($\sim$ 10$^{50}$ erg; \citet{veilleux_2005}). 

\subsection{Driving source of the GS206-17+13 shell}\label{sec:ori1b}

The GS206-17+13 shell is filled with ionized gas that is known as the IC 434 emission nebula. The bright gas emission results from the ionizing flux of the central massive component of the $\sigma$ Ori cluster, the triple-star-system $\sigma$ Ori AB at $d$ = 385 pc \citep{simon_diaz_2011,caballero_2008} that is part of a quintuple system \citep{van_loon_2003, caballero_2007b}, evaporating the L1630 cloud and launches a champagne flow into the expanding bubble. In this respect, it is tempting to attribute the expanding shell to the continuous input of the stellar wind of $\sigma$ Ori AB to its surroundings. However, $\sigma$ Ori is a young system ($t_\m{age}$ = 2-3 Myr; \citealt{caballero_2008}) which exhibits a weak-wind (log($\Mdot$) = -9.7  M$_\odot$ yr$^{-1}$, $v_\m{\infty}$ = 1500 km s$^{-1}$; \citealt{najarro_2011}). This amounts to an integrated mechanical luminosity of $L_\m{mech}$ = 1/2 $\Mdot$ $v_\m{\infty}^2$ $t_\m{age}$ = 1.3 $\times$ 10$^{46}$ erg, of which only $\sim$ 10\% will couple as kinetic energy to the ISM \citep{van_buren_1990,veilleux_2005}. This excludes $\sigma$ Ori as the driving source of the expanding GS206-17+13 shell. 

The Orion OB1b sub-association is centered on Orion's belt and has age estimates ranging from 1.7 Myr \citep{brown_1994} to 8 Myr \citep{blaauw_1964}. However, as noted by \citet{bally_2008}, the supergiant status of the Orion Belt stars ($\zeta$ Ori, $\epsilon$ Ori, and $\delta$ Ori) imply an age of the 1b association of at least 5 Myr. We find 11 stars of spectral type B2 or earlier within the Orion OB1b association \citep{brown_1994}. Remarkably, the massive members of the Orion OB1b association are positioned along the GS206-17+13 shell (Fig. \ref{fig:shell}), which provide clues to their origin.

The GS206-17+13 HI shell is located at $v_\m{LSR}$ = 12.8$\pm$7.8 km s$^{-1}$; \citet{ehlerova_2005}). The most massive star in this $v_\m{LSR}$ range is the B0I supergiant $\epsilon$ Ori at $d$ = 606$^{+227}_{-130}$ pc \citep{van_leeuwen_2007}, with wind parameters log($\Mdot$) = -5.73 M$_\odot$ yr$^{-1}$ \citep{blomme_2002} and $v_\m{\infty}$ = 1600 km s$^{-1}$ \citep{kudritzki_1999}. A simple calculation shows that $\epsilon$ Ori can provide the kinetic energy of the GS206-17+13 shell within $\sim$ 11 Myr (assuming a constant wind strength and 10\% kinetic efficiency), which is a reasonable timescale compared to the estimated age of the Orion OB1b association. Surely, the supergiant stars $\zeta$ Ori and $\delta$ Ori could have been part of the formation of the GS206-17+13 shell. However, the {\em Hipparcos} distances towards $\zeta$ Ori and $\delta$ Ori, $d$ = 225$^{+39}_{-28}$ pc and $d$ = 212$^{+30}_{-23}$ pc \citep{van_leeuwen_2007}, respectively, compared to the average distance of the stars located within the velocity extent of the expanding HI shell ($v_\m{LSR}$ = 12.8$\pm$7.8 km s$^{-1}$), $d$ = 440 pc, renders this scenario implausible. This difference in distance is reflected in the low radial velocity $v_\m{LSR}$ = 1 km s$^{-1}$ for both $\zeta$ Ori and $\delta$ Ori, compared to the velocity extend of the HI shell, which places the stars outside of the bubble. Perhaps, the stars have been thrown out of the cluster through dynamical interaction in the young OB1b cluster. \citet{reynolds_1979} already noted that $\delta$ Ori does not have a clear \HII\ region surrounding it, implying that the ionizing radiation can escape freely into the Orion-Eridanus superbubble cavity \citep{brown_1995} instead of illuminating the surroundings of the star, i.e., the GS206-17+13 shell. We conclude that the stars at $d$ = 440 pc are the best candidates for driving the expansion of the GS206-17+13 shell, with $\epsilon$ Ori as the main contributor.

\subsection{Space velocities of Orion OB1b and approach of $\sigma$ Ori}\label{sec:spaceapproach}

We calculate LSR space velocities of the stars belonging to the Orion 1b association following the method described in \citet{cox_2012}, using proper motions and radial velocities, and correcting for the solar motion \citep{coskonoglu_2011}. Proper motions are taken from \citet{van_leeuwen_2007}, except for the case of $\sigma$ Ori, where due to the complex nature of the system the new reduction of the {\em Hipparcos} catalogue did not produce an acceptable solution ($\mu_{\alpha}$\,cos\,$\delta$ = 22.63$\pm$10.83 mas and $\mu_{\delta}$ = 13.45$\pm$5.09 mas, \citealt{van_leeuwen_2007}, versus $\mu_{\alpha}$\,cos\,$\delta$ = 4.61$\pm$0.88 mas and $\mu_{\delta}$ = -0.40$\pm$0.53 mas, \citealt{perryman_1997}). The new reduction is also inconsistent with proper motion measurements from other members of the central quintuple system of $\sigma$ Ori, $\sigma$ Ori D ($\mu_{\alpha}$\,cos\,$\delta$ = 0.79$\pm$0.92 mas and $\mu_{\delta}$ = -0.90$\pm$0.47 mas, \citealt{van_leeuwen_2007}) and $\sigma$ Ori E ($\mu_{\alpha}$\,cos\,$\delta$ = 2.2$\pm$0.92 mas and $\mu_{\delta}$ = -1.4$\pm$0.47 mas, \citealt{caballero_2007}). Hence, for $\sigma$ Ori the proper motion values of \citet{perryman_1997} are used. Figure \ref{fig:shell} overplots the resulting space velocities for stars with $v_\m{LSR}$ within the velocity extent of the expanding GS206-17+13 HI shell. The plotted trajectories can be traced {\em back} in time, and show that it is possible that the stars have dispersed to their current location projected along the bubble shell. In particular, the stars at the northern side of the GS206-17+13 shell at longitude $l$ $\textgreater$ 206$^{\circ}$ seem to be moving out of the bubble, influencing the morphology of the shell traced at 12 $\mu$m in the direction of movement. This asymmetry is also seen as an elongation of the HI shell in Galactic longitude \citep{ehlerova_2005}, and indicates that the Orion OB1b stars indeed have a relative velocity with respect to the Orion B molecular cloud and the GS206-17+13 shell. We add here as a caveat that due to the the location of Orion in the anti-direction of the Galactic center, the motion of the OB association is mostly directed radially away from the Sun \citep{brown_1994,de_zeeuw_1999}, which complicates precise measurement of proper motions. In addition, the distance towards some of the more distant stars in Orion OB1b gives rise to {\em Hipparcos} parallax errors of 40 - 50\%. In this respect, the Gaia mission \citep{debruijne_2012,perryman_2001} is expected to improve proper motion measurements and distance determinations towards the members of Orion OB1.

Based on its spatial proximity, $\sigma$ Ori has often been classified as being part of the Orion OB1b association. \citet{bally_2008} already noted that the age of the system ($\sim$ 2-3 Myr; \citealt{caballero_2008}) is incompatible with the age of the Orion OB1b association ($\textgreater$ 5 Myr, Sec. \ref{sec:ori1b}). Therefore, \citet{bally_2008} tentatively classified the $\sigma$ Ori cluster as being part of the Orion OB1c association. Following Paper 1, the radial velocity $v_\m{LSR}$ of $\sigma$ Ori and the L1630 cloud are similar ($\sim$ 10 km s$^{-1}$; \citealt{maddalena_1986,gibb_1995}). The proper motion of $\sigma$ Ori translates to a velocity of 10 km s$^{-1}$; we will use this value as the relative velocity {\bf, $v_\m{rel}$,} between the cloud and $\sigma$ Ori AB in Sec. \ref{sec:results}. Then, the trajectory of space motion at 1 Myr, 2 Myr, and 3 Myr shows that $\sigma$ Ori AB can cross the distance from the Orion OB1c region to its current location. Indeed, the sub-association Orion OB1c has an estimated age of 2 Myr to 6 Myr \citep{blaauw_1964,warren_1978,bally_2008}, which would solve the age discrepancy as described above. However, this assumes that the observed space velocity represents the relative velocity between cloud and star, but the actual traversed distance could be smaller if the bulk motion of the Orion region is co-moving with the stars (i.e., $v_\m{rel}$ $\textless$ 10 km s$^{-1}$). In any case, as the $\sigma$ Ori cluster approaches the L1630 cloud, the ionizing flux reaching the cloud surface increases, resulting in the bright photo-evaporation flow seen to date (Sec. \ref{sec:flowparameters}), known as the IC 434 emission nebula.

\section{Dust emission from the IC 434 region}\label{sec:observations}

\begin{figure*}
\centering
\includegraphics[width=18cm]{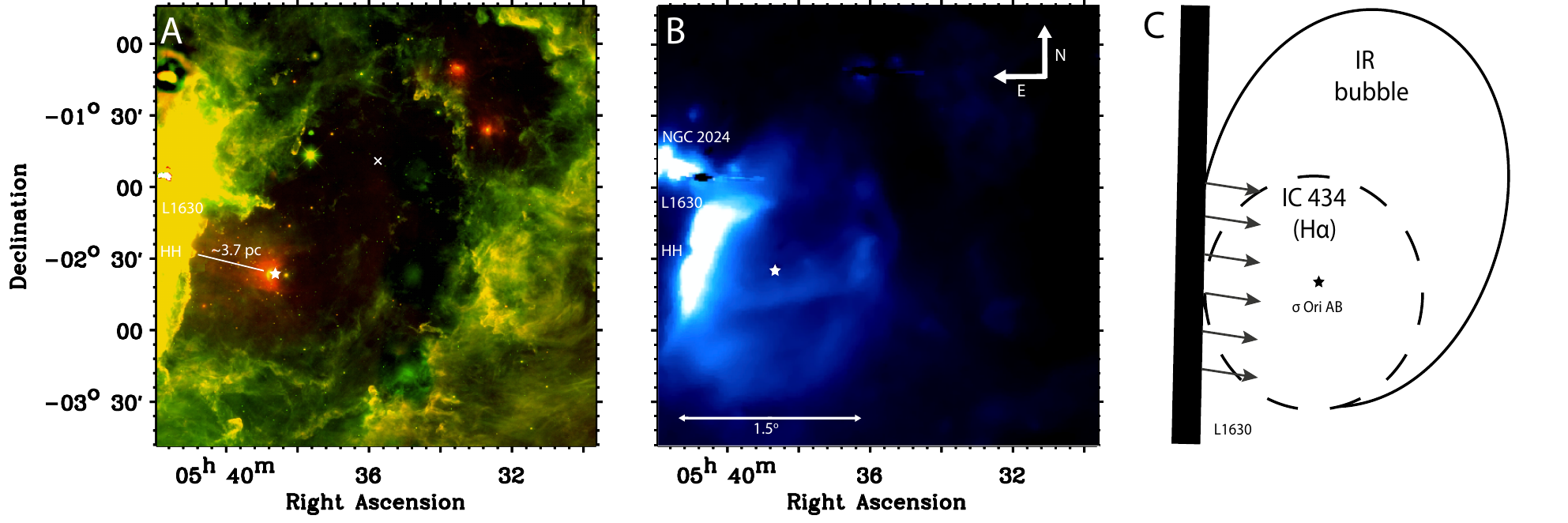} 
\caption{{\bf (A)} Two color composite in the mid-IR of the region around $\sigma$ Ori AB {\em (white asterisk)}. Green is WISE 12 $\mu$m, whereas red is WISE 22 $\mu$m. The large IR counterpart of the GS206-17+13 HI shell surrounds $\sigma$ Ori AB. The east-side of the bubble is marked by the large molecular cloud L1630, which includes the Horsehead Nebula (HH). The projected distance between the Horsehead and $\sigma$ Ori, as shown by the white line along which the Spitzer/IRS observations are positioned (Fig. \ref{fig:llregions}), is about 0.5 degrees or 3.7 pc (at 385 pc; \citealt{caballero_2008}). The white cross reveals the offset position used for the Spitzer/IRS background subtraction. {\bf (B)} Smoothed H$\alpha$ image from SHASSA (4' resolution). The gas emission around $\sigma$ Ori AB (IC 434) originates from a photo-evaporation flow of the L1630 molecular cloud. The bright part appears roughly circular with a radius $\sim$ 0.75$^{\circ}$.  {\bf (C)} Schematic view of the same region, showing the IR bubble {\em (solid outline)} and the H$\alpha$ emission {\em (dashed circle)}, created by the photo-evaporation flow {\em (arrows)} of the L1630 molecular cloud {\em (filled black rectangle)}.} 
\label{fig:ic434}
\end{figure*}

\subsection{Morphological structures}\label{sec:dustdistribution}

Figure \ref{fig:ic434}a shows the mid-IR view of the region surrounding IC 434. The WISE 12 $\mu$m is dominated by PAH emission tracing the PDR surfaces directly adjacent to the IF. However, the gas content in the same region shows a different morphology (Fig. \ref{fig:ic434}b): the bulk of the H$\alpha$ emission concentrates within a smaller region with radius $\sim$ 0.75$^{\circ}$ ($\sim$ 5 pc at a distance of 385 pc) inside the mid-IR bubble (Fig. \ref{fig:ic434}c). The ionized gas originates from the evaporation of the L1630 molecular cloud by the ionizing flux of $\sigma$ Ori AB. This gas flow effectively introduces dust into the ionized gas through momentum transfer.

\begin{figure*}
\centering
\includegraphics[width=18cm]{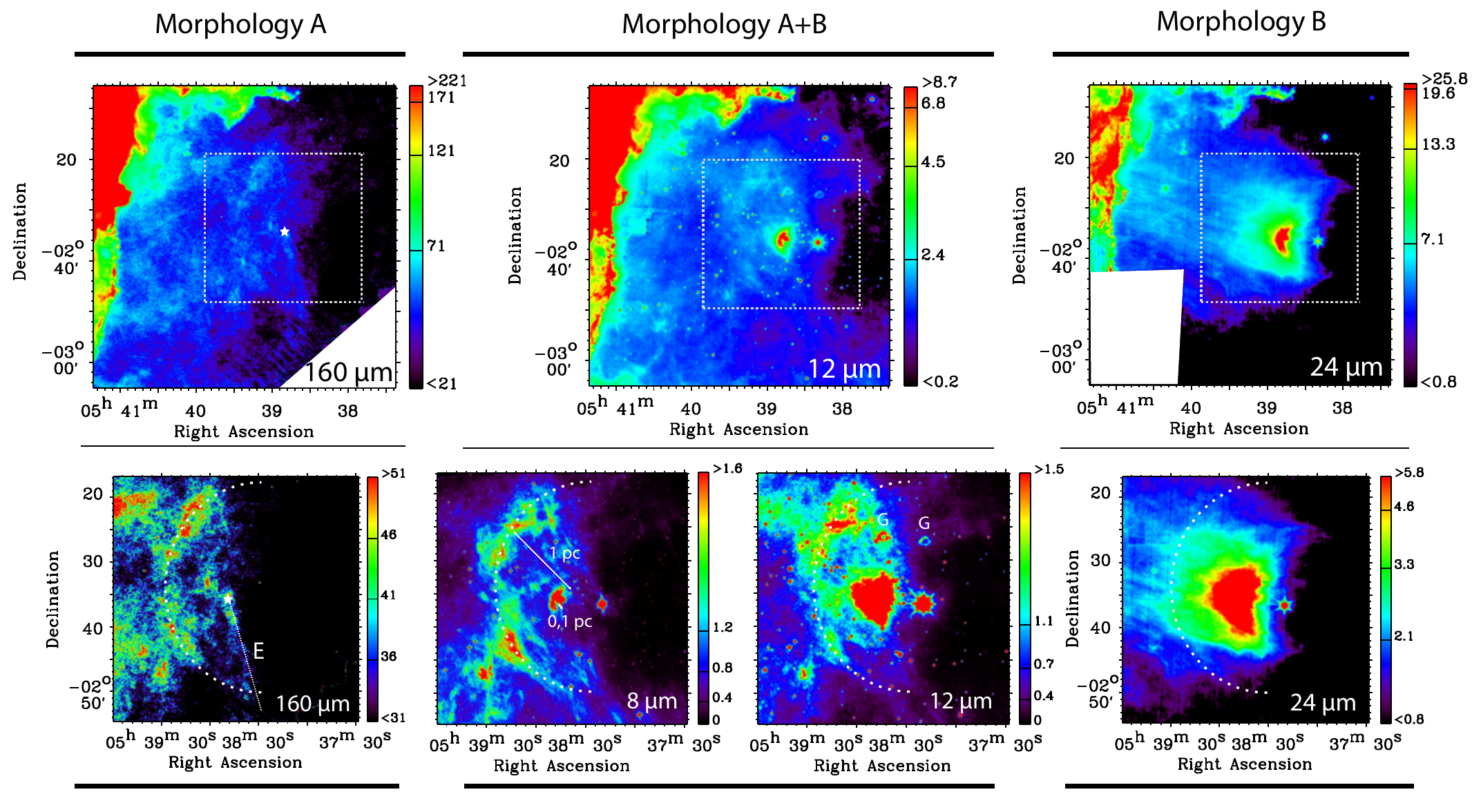} 
\caption{Two different infrared morphologies within IC 434. In all panels: north is up, east is to the left. Morphology A is evident at every wavelength except for 24 $\mu$m, and shows a diffuse structure emanating from the entire L1630 molecular cloud. Morphology B is a relative narrow structure, ranging from the L1630 molecular cloud towards $\sigma$ Ori AB (white asterisk in the 160 $\mu$m images), where it interacts with radiation pressure from the star at projected distance 0.1 $\textless$ $d$ $\lesssim$ 0.4 pc (Paper 1), forming the inner dust wave that is visible at wavelengths $\lambda$ \textless\ 24 $\mu$m. The 8 $\mu$m and 12 $\mu$m images show a combination of both morphologies (A+B). The lower panels zoom into the direct surroundings of $\sigma$ Ori AB and cover the region in the dotted white box seen in the upper panels. The 8 $\mu$m, 12 $\mu$m and 160 $\mu$m images show another arc-like structure at a distance of $\sim$ 1 pc away from the star (dashed line), forming the outer dust wave. The 12 $\mu$m suffer from several ghost features ('G'). The 160 $\mu$m image reveals an extended collimated flow (dotted line) away from the $\sigma$ Ori AB, not seen at 8 $\mu$m and 12 $\mu$m, which is possibly an evaporating member ('E') of the large $\sigma$ Ori cluster. The color bars show the brightness scale in units of MJy/sr.} 
\label{fig:flowstructure}
\end{figure*}

Figure \ref{fig:flowstructure} shows the distribution of dust inside the IC 434 region at wavelengths 8 $\mu$m $\geq$ $\lambda$ $\geq$ 160 $\mu$m. The 24 $\mu$m dust emission surrounding $\sigma$ Ori AB has earlier been attributed to a {\em dust wave} (Paper 1). Here, we report the detection of emission from the ionized gas (Sec. \ref{sec:dustlocation}) at 8 $\mu$m, 12 $\mu$m, and at {\em Herschel} wavelengths longwards of 160 $\mu$m. The 12 $\mu$m WISE filter is broad (ranging from 7 $\mu$m - 17 $\mu$m), and in principle is sensitive to both warm dust radiating at 100 K - 200 K and transiently heated particles, such as Very Small Grains (VSGs) and IR fluorescence from PAHs. While PAHs mainly show up in broad emission features at 5 $\mu$m - 18 $\mu$m, VSGs typically reveal a continuum from $\lambda$ \textgreater\ 13 $\mu$m, which could be a real continuum or a superposition of multiple blended spectral features \citep{cesarsky_2000,abergel_2002}. 

After inspection of the images at different wavelengths, one can distinguish two separate morphological IR emission structures inside the H$\alpha$ emitting volume (Fig. \ref{fig:flowstructure}). On the one hand, there is a component seen at all wavelengths but for 24 $\mu$m (morphology 'A'). This component is characterized through its large diffuse scale and clearly associated with the full extent of the L1630 cloud. The morphology is spread throughout the entire IC 434 nebula, ultimately culminating in an arc-shaped increase in emission to the east of $\sigma$ Ori AB at a projected distance of $d$ $\sim$ 1 pc (best observed at 8 $\mu$m), together with a clear drop-off in emission to the west of the star. For the remainder of this work, we will refer to this arc-like structure at $d$ = 1 pc as the outer dust wave (ODW), due to the striking resemblance of the structure with the dust wave observed at 0.1 pc (see below). On the other hand, there is the emission detected with WISE and Spitzer at 22 $\mu$m and 24 $\mu$m which is clearly distinct from morphology A, peaking at $d$ $\sim$ 0.1 pc and revealing a relative narrow flow (morphology 'B'). This flow seems to be connected to a bright part of the L1630 cloud which is almost symmetrically positioned around the Horsehead Nebula, again forming an arc-like structure that peaks in intensity at $d$ $\sim$ 0.1 pc. We will refer to this structure as the inner dust wave (IDW). Close to the star, the IDW lights up at 12 $\mu$m and 8 $\mu$m due to heating of the grains and an increased dust-to-gas ratio (Paper 1).

\begin{figure*}
\centering
\includegraphics[width=16.5cm]{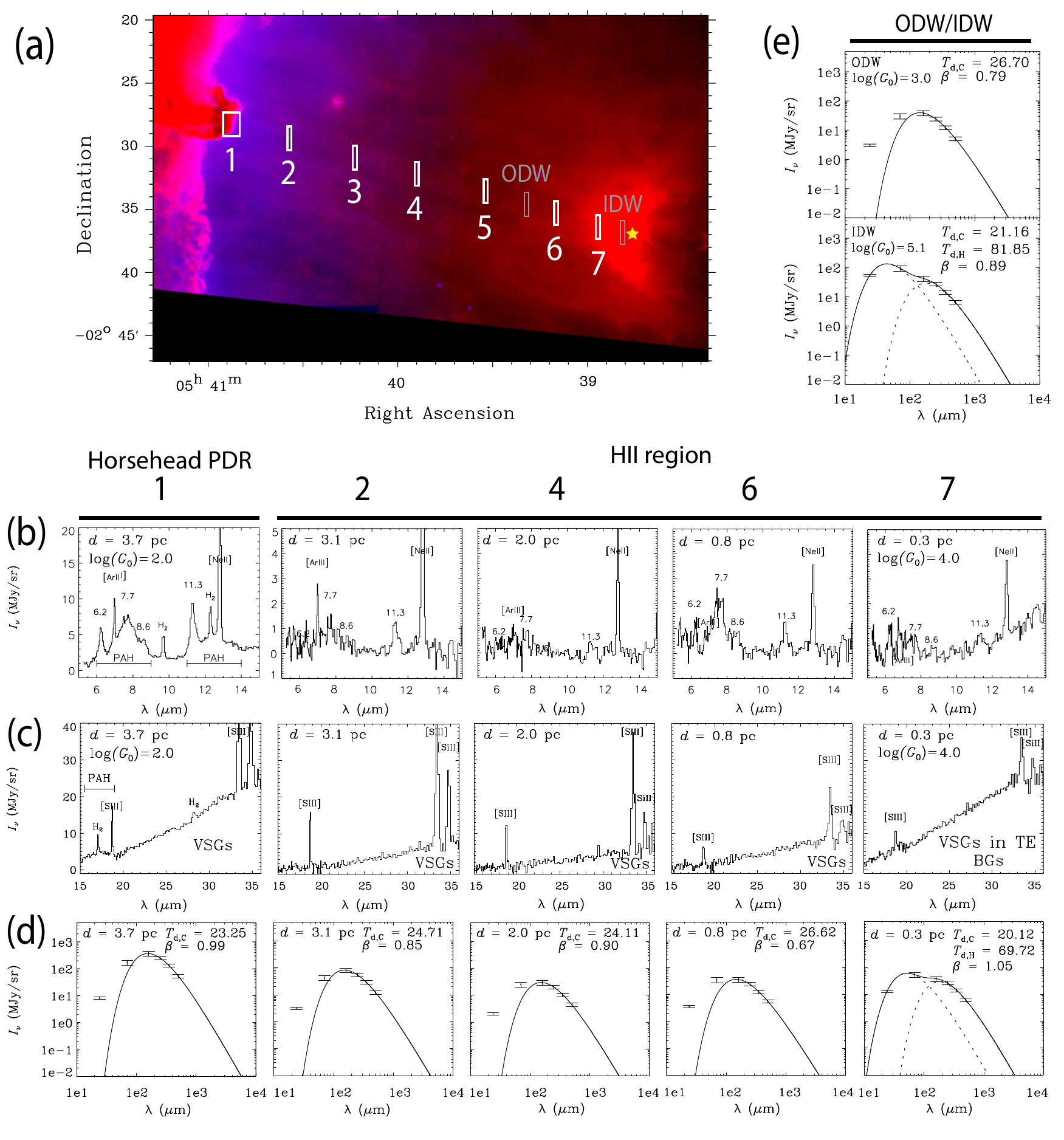} 
\caption{{\bf(a)} The mid-IR view of the Horsehead - $\sigma$ Ori AB {\em (yellow asterisk)} region. Red is MIPS 24 $\mu$m, blue is H$\alpha$ from the KPNO 4.0m Mayall telescope. Overplotted in white rectangles are the regions for which Spitzer IRS/spectra and SEDs were extracted. The gray rectangles are the location of the peak intensities of the outer dust wave (ODW) and inner dust wave (IDW) for which only the SED is available. {\bf (b)} From left to right, Spitzer/IRS SL spectra at a distance $d$ = 3.7 pc (located partly inside the Horsehead PDR), 3.1 pc, 2.0 pc, 0.8 pc (inside the \HII\ region) and 0.3 pc (projected on top of IDW) from $\sigma$ Ori AB are extracted, where $d$ is the distance along a straight line connecting $\sigma$ Ori AB and the Horsehead. Indicated are gas emission lines, the PAH plateau at 6 - 9 $\mu$m and at 11 - 14 $\mu$m, and the incident radiation field $G_\m{0}$. PAH emission is clearly observed throughout the \HII\ region. {\bf (c)} Same as (b), but for the IRS LL modules. Labelled beneath the curve is the source of the rising continuum in the mid-IR. {\bf (d)} Far-infrared spectral energy distribution of the regions. In all cases except for region 7, a single-component fit is conducted on the SED, as VSGs emit significantly at $\lambda$ $\leq$ 70 $\mu$m. In region 7, we assume thermal equilibrium for the warm component and add a second component to the fit. {\bf(e)} Spectral energy distributions for the ODW and IDW.}
\label{fig:llregions}
\end{figure*}

\subsection{Spectral energy distributions}\label{sec:obsproperties}

To get insight in the properties of the emitting dust in the \HII\ region, we constructed SEDs from the WISE, MIPS, PACS and SPIRE photometric points. All images were convolved to the point spread function of the SPIRE 500 $\mu$m band using the kernels described in \citet{aniano_2011}. Subsequently, a common 'dark' spot in the maps was chosen for background subtraction (centered at 05$^h$37$^m$00$^s$ -01$^d$52$^m$29$^s$). The absolute uncertainties on the photometry are estimated at 10\% for the WISE and Spitzer images, while we use uncertainties of 20\% for PACS and 15\% for SPIRE, respectively \citep{bernard_2010}. We extract SEDs corresponding to the regions covered by IRS (regions 1-7; Fig. \ref{fig:llregions}) and complement these with two regions at both dust arcs, located at 1 pc (ODW) and 0.1 pc (IDW), respectively. The results are shown in Fig. \ref{fig:llregions}. 

\subsection{Are we tracing emission from the ionized gas?}\label{sec:dustlocation}

The analysis of the remainder of this paper assumes that the diffuse dust emission from  morphology A and B (Sec. \ref{sec:dustdistribution}) originates from within the ionized gas of the IC 434 emission nebula. For morphology B, this is obvious from the bright interaction zone close to $\sigma$ Ori AB (Paper 1). Below, we will argue that the same is true for morphology A. 

Both for the continuum and the spectroscopic observations, we have subtracted a background value to exclude dust associated with the GS206-17+13 shell and contamination along the line-of-sight (Fig. \ref{fig:ic434}). This method may leave some residual structure if the bubble shell is not homogeneous. However, H$_\m{2}$ emission at 9.7 $\mu$m, 12.3 $\mu$m, and 17.0 $\mu$m immediately dissapear (Fig. \ref{fig:llregions}b\&c) when moving from region 1 (dominated by the Horsehead PDR) towards region 2 - 7, indicating a transition from the shielded PDR to the atomic/ionized gas, attesting to the success of the correction for the background emission from any surrounding neutral material. Moreover, morphology A is clearly separated from the L1630 molecular cloud, filling the H$\alpha$ emitting region (Fig. \ref{fig:llregions}) with diffuse emission, ranging from the L1630 molecular cloud to $\sigma$ Ori. Furthermore, morphology A contains a limb-brightened arc at a projected distance of 1 pc east of the star (more pronounced in associated 8 $\mu$m emission; see below), upstream in the flow of ionized gas, and it abruptly terminates when moving towards the west-side of the star, both indicative of an interaction with $\sigma$ Ori AB, and arguing against a connection to background material. Indeed, in the photo-evaporation flow, most of the material is expected to be located near the symmetry axis connecting the star and the cloud (Sec. \ref{sec:flowparameters}). Within this framework, the total diffuse dust emission located between the star and the cloud represents a dust reservoir that has evaporated of the L1630 molecular cloud, and is moving along with the ionized gas, some of which has not yet reached the interaction zone ahead of $\sigma$ Ori AB. 

\begin{figure}
\centering
\includegraphics[width=9cm]{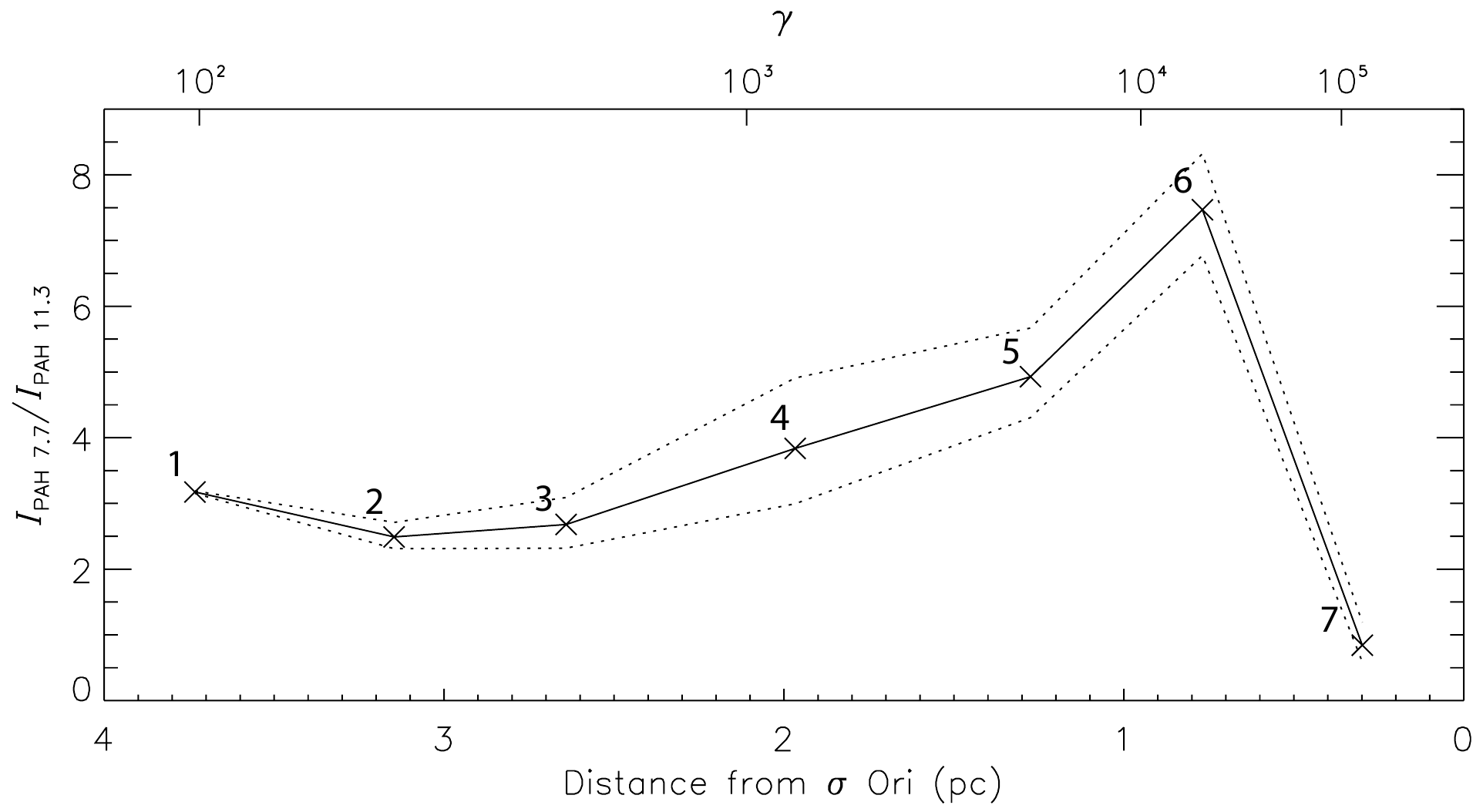} 
\caption{The evolution of the PAH 7.7 $\mu$m over 11.3 $\mu$m intensity ratio as a function of distance into the \HII\ region, tracing the ionization state of the PAH molecules \cite[e.g.][]{bauschlicher_2002, szczepanski_1993}. The intensities of the bands are measured with PAHFIT \citep{smith_2007}. The ratio shows a systematic rise with distance into the \HII\ region, indicating that the molecules reside within the ionized gas, and a dramatic drop in the region closest to the ionizing star. The numbers correspond to the regions drawn in Fig. \ref{fig:llregions}; the dotted lines show the 1$\sigma$ uncertainty constrained by the quality of the spectra. The upper axis plots the ionization parameter $\gamma$ \citep{tielens_2005}.}
\label{fig:pahevolution}
\end{figure}

Emission observed in the WISE 12 $\mu$m band and IRAC 8 $\mu$m closely follow the spatial morphology of the cold dust of Morphology A. As can be seen from the Spitzer/IRS spectra shown in Fig. \ref{fig:llregions}b\&c, the WISE 12 $\mu$m is picking up PAH emission from the 7.7 $\mu$m and 11.3 $\mu$m band. The dust continuum from VSGs inside the \HII\ region (region 2 - 6) only rises at $\lambda$ $\textgreater$ 20 $\mu$m (see Sec. \ref{sec:vsgs}), well outside the WISE filter edge ($\sim$ 17 $\mu$m). We note that emission from [NeII], arising from the IC 434 emission nebula, potentially contaminates the detection in the broad WISE filter. However, if the [NeII] emission were to dominate the observed flux in the WISE band, this would follow the smooth, exponential appearance from the H$\alpha$ emission measure (Fig. \ref{fig:ic434}b), which is not seen in the observations. We conclude that WISE traces PAH emission. The IRAC 8 um band also traces PAH emission. By realizing that the cold dust observed by {\em Herschel}, and the PAH emission as seen by {\em WISE} and {\em Spitzer} are spatially related (Fig. \ref{fig:llregions}), we can use the evolution of the mid-IR PAH spectrum to answer the question if morphology A traces a dust component that is located {\em inside} the ionized gas (Fig. \ref{fig:pahevolution}). In this respect, we note that \citet{compiegne_2007} already showed that the PAH emission originates from within the ionized gas through correlation with ionized gas tracers.

The intensity ratio of the 7.7 $\mu$m to 11.3 $\mu$m PAH features is a well-known indicator of the charge state of PAHs (e.g., \citealt{szczepanski_1993,bauschlicher_2002}), as neutral species emit significantly less in the 6 - 9 $\mu$m region compared to charged species \citep{allamandola_1999}. Figure \ref{fig:pahevolution} plots the 7.7/11.3 $\mu$m intensity ratio, which initially decreases from region 1 to 2: \citet{compiegne_2007} argue that, initially, the PAHs become more neutral due to the increase of free electrons across the IF. Subsequently, from region 2 to region 6 the 7.7/11.3 $\mu$m ratio shows a steady increase inside the \HII\ region, indicating an overall increase of the (cationic) charge state of the PAHs. Thus, the PAHs are located inside the ionized gas: we are probing emission from regions that are located increasingly closer to the star. The combination of a decrease of electrons available downstream in the ionized flow with a gain of stellar photons increases the ionization parameter $\gamma$ \citep{tielens_2005}, defined as $G_\m{0}T^\m{1/2}/n_\m{e}$. Here, $G_\m{0}$ is the hardness of the radiation field in units of the Habing field \citep{habing_1968}, $T$ is the temperature of the (ionized) gas, and $n_\m{e}$ is the electron number density. A 'typical-sized' interstellar PAH with 50 carbon atoms will become more positively charged as $\gamma$ increases, and the cationic state will dominate at $\gamma$ $\gtrsim$ 10$^4$ \citep{tielens_2005}. In region 7, we are tracing a region PAH-free cavity around the star  and we cannot trace the charge ratio reliably anymore (see Sec. \ref{sec:pahs}). The observed relation of the 7.7/11.3 $\mu$m ratio with $\gamma$ (Fig. \ref{fig:pahevolution}) is in reasonable agreement with observations of the Orion Bar and NGC 2023 \citep{galliano_2008}.

Another proxy for the charge state of PAHs is the 11.0 $\mu$m feature due to C-H out of plane bending modes of cationic PAHs \citep[e.g.,][]{hony_2001,rosenberg_2011}. Indeed, we find a hint of the 11.0 $\mu$m feature in regions 5 - 6 close to the star, but the quality of the IRS spectrum prevents us from making a conclusive statement on a detection of the 11.0 $\mu$m satellite feature. 

In summary, we argue that morphology A represents a dust population entrained within ionized gas of the IC 434 emission nebula because of the following:

\begin{enumerate}
\item[(1)] We subtracted the background to exclude dust associated with the GS206-17+13 shell: analysis of the Spitzer/IRS spectrum reveals the absence of the H$_\m{2}$ feature immediately after leaving the Horsehead PDR: there is no neutral region associated with the spectra of regions 2 - 7 (Fig. \ref{fig:llregions}).
\item[(2)] PAH emission shows a tight correlation with the long wavelength dust emission associated with morphology A (Fig. \ref{fig:flowstructure}).
\item[(3)] PAHs are increasingly charged when moving to regions projected closer towards the $\sigma$ Ori AB. Thus, the PAHs are located {\em inside} the ionized gas. This conclusion was also reached by \citet{compiegne_2007}, based upon observations of the Horsehead nebula. Consequently, the same holds for the cold dust emission from morphology A that is seen co-spatial with the PAH emission.
\item[(4)] The diffuse emission from morphology A culminates in an arc-shaped emission around $\sigma$ Ori AB, accompanied by a clear-drop in emission and a cavity behind the star, both indicating an interaction of the dust with $\sigma$ Ori AB close to the symmetry axis connecting the star and cloud (Sec. \ref{sec:introducingdust}).
\end{enumerate}

\subsubsection{VSGs: 15 $\mu$m - 35 $\mu$m mid-infrared spectra}\label{sec:vsgs}
The IRS observations were designed to follow the evolution of the mid-IR spectrum in an increasing radiation field. The 15 $\mu$m - 35 $\mu$m spectra (Fig. \ref{fig:llregions}c) show gas emission lines from [S III] and [Si II], typical for an ionized nebula. In addition, as region 1 is (partly) located inside the Horsehead PDR, molecular H$_\m{2}$ emission is seen along with the PAH emission plateau, ranging from 16 $\mu$m - 19 $\mu$m. A steep continuum at $\lambda$ $\textgreater$ 15 $\mu$m reveals the presence of VSGs, as the intensity of the radiation field, log($G_\m{0}$) = 2.0, is too low for classical grains, or Big Grains (BGs), to produce continuum emission in the mid-IR. This continuum decreases between region 1 and region 2, but persists throughout the \HII\ region and shows a remarkable constant spectral slope (see Tab. \ref{tab:color}) as revealed by the flux continuum ratio at 20 $\mu$m and 30 $\mu$m, [20/30]. The carbonaceous nature of the carriers of the mid-IR continuum emission (often assumed for the VSGs in dust models) is confirmed through the absence of the silicate emission feature at 20 $\mu$m \citep[e.g.,][]{van_boekel_2003}, but for an opposing view see \citet{li_2002}.

The constant slope strongly implies the presence of stochastically-heated VSGs that survive throughout the \HII\ region, as a constant flux ratio does not reflect a dust population in TE inside an increasing radiation field. The decrease in observed flux between the PDR (region 1) and the \HII\ region (region 2 - 7) likely comprises a sub-population of VSGs that do not survive the transition from the PDR into the \HII\ region. We have labelled the responsible carriers in Fig. \ref{fig:llregions}b: note that in region 7, where log($G_\m{0}$) = 4.0, both VSGs and BGs are expected to be in TE and can both be responsible for the increase of the ratio [20/30] in region 7. In this respect, the observed dust temperature of this component at the IDW (73-82 K), and the reduced UV-to-IR opacity (Sec. \ref{sec:opacity}) argue in favor of BGs, consistent with the findings in Paper 1. Note that these BGs can both be of carbonaceous or silicate nature; the peak temperature is too low for the 10 $\mu$m and 20 $\mu$m silicate features to be in emission.

\begin{table}
\centering
\begin{tabular}{l|c|c|c|c|c|c|c}\hline \hline
Region & 1 & 2 & 3 & 4 & 5 & 6 & 7 \\
Location & PDR & \HII\ &\HII\ &\HII\ & \HII\ & \HII\ & DW \\ \hline
log($G_\m{0}$) & 2.0 & 2.2 & 2.3 & 2.5 & 2.9 & 3.3 & 4.2 \\
$[20/30]$  & 0.28 & 0.20 & 0.21 & 0.19 & 0.22 & 0.21 & 0.33 \\ \hline \hline
\end{tabular}
\caption{The strength of the radiation field $G_\m{0}$ \citep{habing_1968}, and the [20/30] continuum flux ratio for the regions in Fig. \ref{fig:llregions}. Region 1 is located partly inside the Horsehead photo-dissociation region (PDR), whereas regions 2-7 are located inside the \HII\ region. Region 7 is projected on top of the dust wave surrounding $\sigma$ Ori AB. The [20/30] ratio is insensitive to the increase of $G_\m{0}$ inside the \HII\ region, except in region 7, where the dust reaches thermal equilibrium (see text).}
\label{tab:color}
\end{table}

\subsubsection{Cold dust: far-infrared spectral energy distribution}\label{sec:colddust}

It is common to model the observed dust SED with a single modified blackbody (MBB) function. However, due to the presence of VSGs in the \HII\ region, which contribute to the emission observed at $\lambda$ $\leq$ 70 $\mu$m, this is not a straightforward task \citep{anderson_2012}. As the SED clearly reveals the presence of a warm dust component (Fig. \ref{fig:llregions}), we exclude the 24 $\mu$m and 70 $\mu$m data for all regions, except for region 7 and the IDW. Here, we assume that the dust component emitting at 24 $\mu$m an 70 $\mu$m has reached TE with the radiation field, and perform a two-component blackbody fit to the SED. 

The cold component reveals a striking constant appearance throughout the \HII\ region, characterized by temperatures of $T_\m{d,C}$ $\sim$ 23 K - 26 K, and low spectral emissivity, $\beta$ $\sim$ 0.7 - 1.0. Still, these values of $\beta$ are within what is expected for the derived dust temperatures \citep{veneziani_2010, planck_col_2011}. Note however, that the parameters $T_\m{d}$ and $\beta$ seem degenerate, and derived values are sensitive to systematic effects induced from photometric calibration uncertainties to different grain temperatures superpositioned along the line-of-sight \citep[see][]{juvela_2012}, and dust optical properties \citep{jones_2014}. Nevertheless, the constant appearance of the SED is curious, considering that the impinging radiation field varies significantly (Tab. \ref{tab:color}). For the warm component, we fix $\beta$ = 1 due to the lack of data points with respect to free parameters, a value appropriate for the spectral index in the (mid-)IR \citep{mennella_1995}. Then, the warm component reaches a maximum temperature of $T_\m{d,H}$ = 82 K at the position of the IDW ($d$ = 0.1 pc). 

As the cold dust does not contribute much at 70 $\mu$m at the location of the IDW ($\textless$ 10\%), the addition of the extra component at long wavelengths does not change the results obtained in Paper 1. However, the peak dust temperature of the hot dust depends on the choice of $\beta$: for a value of $\beta$ = 1.8 for the hot component (as was used in Paper 1), the peak temperature decreases to $T_\m{d,H}$ = 73 K. The addition of the hot component does affect the fit in region 7 by increasing $\beta$ and lowering $T_\m{d,C}$, because here the hot component contributes significantly at 70 $\mu$m. This implies that the single-component fits in region 1 - 6 overestimate $T_\m{d,C}$ slightly. Indeed, by fixing the spectral emissivity at $\beta$ = 1.5 (as measured by {\em Planck} for IC 434 \citet{abergel_2013}), yield $T_\m{d,C}$ = 19 K - 20 K for region 1 - 6. The larger value of $\beta$ as measured by {\em Planck} is most likely caused by different dust populations and mixing of temperatures, as the Planck model samples the spectral emissivity at a 30' resolution, and does not correct for contamination along the line-of-sight. The high angular resolution of {\em Herschel}, and our careful background subtraction isolates the emission from the ionized gas. Still, to determine the actual temperature and spectral emissivity of dust in the \HII\ region with a single MBB fit is not straightforward, due to the presence of multiple dust components {\em inside} the ionized gas, as reflected by the SEDs.

\begin{table}
\centering
\begin{tabular}{l|l|c|c|c}\hline \hline
&  & Reg. 1 & ODW & IDW \\ \hline
\multirow{ 2}{*}{} & $d$ (pc) & 3.7 & 1 & 0.1 \\ 
& log($G_\m{0}$) & 2.0 & 3.0 & 5.1 \\ \hline \hline
MBB & $T_\m{d,C}$ (K) & 23 (19) & 27 (20) & 82 (74) \\ \hline
\multirow{ 4}{*}{DUSTEM} & $T_\m{eq,S}$ (K) 0.1 $\mu$m & 31 & 47 & 100 \\
& $T_\m{eq,S}$ (K) 1.0 $\mu$m & 20 & 31 & 65 \\
& $T_\m{eq,C}$ (K) 0.1 $\mu$m & 34 & 55 & 130  \\
& $T_\m{eq,C}$ (K) 1.0 $\mu$m & 21 & 33 & 74 \\ \hline
\end{tabular}
\caption{Dust temperatures $T_\m{d}$ for grains at the location of region 1, the outer dust wave (ODW), and the inner dust wave (IDW), derived through 1) a modified-blackbody-fit (MBB) and 2) DUSTEM modeling of the equilibrium temperature for a 0.1 $\mu$m silicate grain and 1.0 $\mu$m silicate grain ($T_\m{eq,S}$) and amorphous carbon grain ($T_\m{eq,C}$). Also listed are the projected distance $d$, and the hardness of the radiation field, $G_\m{0}$ \citep{habing_1968}. The observed cold component $T_\m{d,C}$ is calculated with $\beta$ as a free parameter (see Fig. \ref{fig:llregions}), and by choosing $\beta$ = 1.5 as given by the Planck dust model \citep{abergel_2013}, denoted between brackets.}
\label{tab:dusttemp}
\end{table}

In Sec. \ref{sec:dustdistribution}, we showed that the far-IR emission in IC 434 is dominated by morphology A, which fills the \HII\ region and forms the ODW at $d$ = 1 pc to the east of the star. The dust population accountable for morphology A is distributed {\em inside} the \HII\ region, ranging from region 1 ($d$ = 3.7 pc) to the ODW ($d$ = 1 pc) (Sec. \ref{sec:dustlocation}), close to the symmetry axis (Sec. \ref{sec:introducingdust}). Radiation pressure from the star prohibits the dust from drawing closer (Sec. \ref{sec:dustwave}). In a first attempt to constrain the properties of the grains responsible for morphology A, we use DUSTEM \citep{compiegne_2011} to predict the shape of the dust FIR SED by using a radiation field evaluated at the projected distance to $\sigma$ Ori AB. Compared to the single-temperature MBB fit, DUSTEM allows for a range of equilibrium temperatures, a natural consequence of the size distribution and composition of interstellar dust. 

\begin{figure}
\centering
\includegraphics[width=9cm]{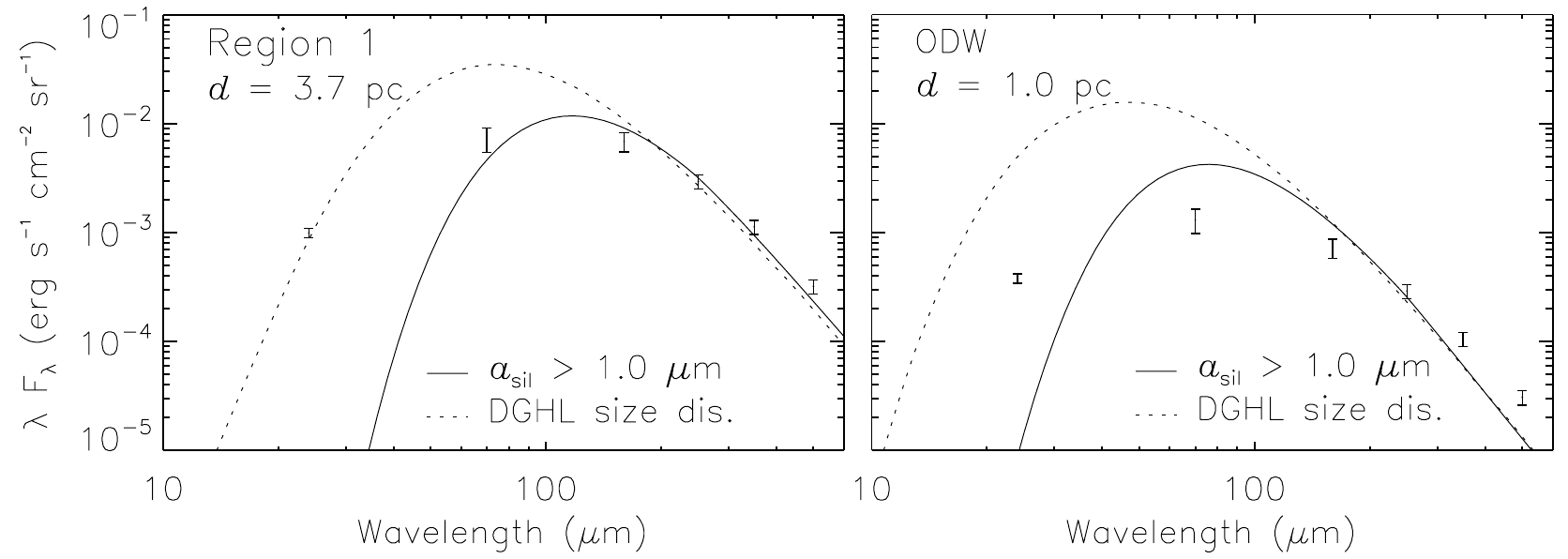} 
\caption{Modeling of the far-infared spectral energy distribution using DUSTEM \citep{compiegne_2011}. Two panels are shown, corresponding to two different regions (Fig. \ref{fig:llregions}). The left panel shows region 1 ($d$ = 3.7 pc); the right panel displays the outer dust wave region (ODW; $d$ =  1 pc). Overplotted is the predicted dust emission for two different composition and size distribution of grains (see text).}
\label{fig:sed_dustem}
\end{figure}

Figure \ref{fig:sed_dustem} shows the predicted SED by DUSTEM for both region 1 and the ODW. The result are shown for two scenarios. First, we use grain species (both amorphous carbon and silicates) and size distributions that successfully reproduce the extinction and emission properties of a selected region of the diffuse ISM at high Galactic latitude (the DHGL model; \citealt{compiegne_2011}). In this case, the small end of the size distribution dominates in surface area, which causes the predicted dust emission to peak at short wavelengths, causing DUSTEM to be unable to reproduce the observed SED. Second, by {\em only} including very large silicate grains in the size distribution a reasonable fit can be obtained, as explained below. Table \ref{tab:dusttemp} compares $T_\m{d,C}$ from the modified-blackbody fit, at the location of region 1 and the ODW, with equilibrium dust temperatures of graphite and silicate grains calculated with DUSTEM. The results for the IDW region are also listed. While DUSTEM predicts that the grains are significantly heated, resulting in a total temperature increase of $\sim$ 50\% between region 1 and the ODW (because of the increase of stellar flux as they approach the star), the modified black body fits reveal that the cold dust component only increases $\sim$ 15\% in temperature. This result highlights the main puzzle presented by the FIR SEDs of the \HII\ region: a cold dust population, whose temperature is rather insensitive to an increase in radiation field. Below, we will outline two scenarios that might explain the remarkable properties of the grains responsible for Morphology A.

A reasonable fit for region 1 can be obtained by {\em only} including silicate grains larger than 1 micron to the fit. Silicates have lower equilibrium temperatures compared to amorphous carbon, and the size distribution composed of very large grains pushes the peak of the predicted dust emission to longer wavelengths. Still, this grain population is not able to fit the observed SED at the ODW region, as the predicted dust emission peaks blueward of the observed SED. This can perhaps be solved if the grains deviate from the classical spherical description. For example, the equilibrium temperature $T_\m{eq}$ of fluffy aggregates is thought to be lower, compared to their compact counterparts (typically about 10\% - 20\%; \citealt{fogel_1998}). This is because UV, visible and near-IR absorptivity remain similar as fluffiness is increased, whereas the sub-mm emissivity increases significantly \citep{bazell_1990}. An intuitive explanation for the presence of large aggregates inside the \HII\ region arises when realizing that the grains are 'freshly' evaporated from the molecular cloud (Sec. \ref{sec:introducingdust}) in which the grain are thought to coagulate to large, fluffy aggregate structures \citep{ossenkopf_1994,ormel_2009}. This argumentation connects well with the work of \citet{martin_2012}, that used BLAST and IRAS images of selected regions in Vela of moderate column density to show considerable variations in, e.g., the submillimeter opacity (as compared to the diffuse ISM). The authors argued that the inferred grain properties may reflect past histories in the evolution of the grains, such as coagulation in dense regions.

Besides a rather rigorous change in the size distribution and invoking fluffiness or porosity, another explanation for the observed low dust temperatures for Morphology A (as compared to predictions made by DUSTEM) can be linked to recent results from, e.g., the {\em Planck} mission. \citet{abergel_2013} argue that $T_\m{obs}$ is not a simple tracer of the radiation field, but rather probes variations in dust properties (grain structure, material change, and size distributions). Similar results are reported in \citet{aniano_2014}: through comparison with the \citet{draine_2007} physical model, the authors show that the location of the peak wavelength of the dust SED does not only trace the intensity of the radiation field (like DUSTEM), but also intrinsic variations of dust properties. Other results from the {\em Planck} mission \citep{abergel_2013} include that dust in specific region of the ISM is more emissive than those used in dust models from \citet{li_2001} and \citet{compiegne_2007}, which would lower the equilibrium temperature with respect to that computed using these grain models. In conclusion and as pointed out by \citet{jones_2014}, dust models that do not allow for the evolution of grain properties as they transit between different regions of the ISM, will be challenged in the light of new observational evidence obtained by {\em Planck} and {\em Herschel}.

\subsubsection{Dust- and gas mass inside IC 434}\label{sec:mass}

The total dust mass in IC 434 is determined by deriving the average $\tau_\m{353}$ from the Planck model R1.20 \citep{abergel_2013} inside an aperture that encloses the bright H$\alpha$ emission from the emission nebula. We find $\tau_\m{353}$ = 1.5 $\times$ 10$^{-5}$. The total dust mass can then be estimated directly through $M_\m{d,IR}$ = $\frac{\tau_{\m{\lambda}}}{\kappa_\m{\lambda}} S$ \citep{hildebrand_1983}, where $S$ is the size of the aperture and $\kappa_\m{\lambda}$ is the grain opacity at wavelength $\lambda$. If $\kappa_\m{353}$ = 1.9 cm$^{2}$ g$^{-1}$ \citep{weingartner_2001}, then the total dust mass in IC 434 equals $M_\m{d,IR}$ = 2.3 $M_\odot$.

The total gas mass inside IC 434 can be estimated by integrating the gas density profile, given by the H$\alpha$ observations (Sec \ref{sec:flowparameters}), out to a distance of 9 pc (the apparent extent of the bright H$\alpha$ emission). This yields a total hydrogen column of $N_\m{H}$ = 1.6 $\times$ 10$^{20}$ cm$^{-2}$. We assume that the gas evaporates of a 9 pc $\times$ 9 pc surface from the L1630 molecular cloud (Sec. \ref{sec:flowparameters}). Then, the total gas mass in the IC 434 emission nebula is $M_\m{g}$ $\approx$ 100 $M_\odot$. The total dust-to-gas ratio in the IC 434 region is therefore $\sim$ 0.02, similar to that seen in the diffuse ISM ($\sim$ 0.01). A thorough study of gas-to-dust ratios in large-scale photo-evaporation flows could reveal if dust entrainment significantly enriches the ISM with grain materials.

\subsubsection{Dust opacities of the ODW in the UV and IR}\label{sec:opacity}

Following the method and using parameters described in Paper 1 for the IDW, we estimate the optical depth of the ODW at UV wavelengths through aperture photometry at $\tau_\m{uv}$ = 1.2 $\times$ 10$^{-3}$, with a corresponding dust mass of  $M_\m{d,UV}$ = 7.4 $\times$ 10$^{-3}$ $M_\odot$. In contrast, following the same method for $M_\m{d,IR}$ as described in Sec. \ref{sec:mass} and by using $\kappa_\m{160}$ = 10 cm$^{2}$ g$^{-1}$ \citep{weingartner_2001}, then the total mass in the ODW from the IR observations at 160 $\mu$m is $M_\m{d,IR}$ = 0.16 $M_\odot$. Thus, we find for the dust content in the ODW, $M_\m{d,IR}$/$M_\m{d,UV}$ = 22. This effect was already noted for the warm component in Paper 1 and in the W3 region \citep{salgado_2012}, and implies that the dust opacities at UV and IR wavelengths for the cold dust component are also very different from that seen for the diffuse ISM: this may well be a general characteristic for dust inside \HII\ regions. Similar results are observed in (molecular) clouds where the dust opacity is altered, reflected in a flattening of the dust extinction curve towards UV wavelengths \citep{cardelli_1989}. Often, this is attributed to dust coagulation or a dearth of small particles in the dust size distribution. 

\subsubsection{Summary on the observations}

The total dust-to-gas ratio inside IC 434 is $\sim$ 0.02. The ODW contains 5 - 10\% of the total dust content (2.3 $M_\odot$) inside IC 434. The dust content inside the ionized gas can be divided in two different populations, following two separate morphologies (Fig. \ref{fig:llregions}).

Morphology B (Fig. \ref{fig:ic434}) shows the signatures of carbonaceous VSGs in regions 1 - 6, responsible for the constant [20/30] flux inside the \HII\ region. The emission from this component is most pronounced at 24 $\mu$m. The reduced UV-to-NIR opacity, and the inferred peak dust temperature (73 K - 82 K) at $d$ = 0.1 pc indicate that large grains also populate morphology B and perhaps dominate the emission from the IDW (results from Paper 1). These BGs are only visible close to the star as the grains heats up.

Morphology A (Fig. \ref{fig:ic434}) is a combination of cold dust at $\lambda$ $\geq$ 160 $\mu$m, and mid-IR emission from PAHs at 8 $\mu$m and 12 $\mu$m emission that is tightly correlated with the cold dust throughout the \HII\ region. The PAHs show an increasing level of charge state in regions at smaller projected distances from the ionizing star, providing convincing evidence that the PAHs are located inside the ionized gas and therefore, the grains from morphology A are located in the \HII\ region as well.

Dust temperatures of morphology A are between 23 K $\textless$ $T_\m{d,C}$ $\textless$ 26 K (depending on the exact value of $\beta$); this is colder than expected for a typical size distribution of dust in the diffuse ISM, as predicted by DUSTEM. We have pointed out that the cold temperature can be explained by (porous/fluffy) micron-sized grains. This is consistent with the reduced UV-to-IR opacity compared to that observed for dust in the diffuse ISM, and we have connected these observations to the molecular cloud phase from which the grain are evaporated. However, even these large grains cannot explain the temperature dependence that seem insensitive to an increase in intensity of the radiation field. This might confirm recent results, indicating that the peak wavelength of the SED is not determined by the intensity of the radiation field alone \citep{abergel_2013,aniano_2014}, but also traces changes in grain properties \citep[e.g.,][]{jones_2014}.

\section{The flow of dust into the IC 434 region}\label{sec:introducingdust}

As has been discussed in Sec. \ref{sec:spaceapproach}, $\sigma$ Ori approaches the L1630 molecular cloud to create a champagne flow that enters into the expanding bubble, blown by the Orion OB1b association (Fig. \ref{fig:shell}). The increase in the ionizing flux impinging the L1630 molecular cloud with time will strengthen the champagne flow. Note that this scenario differs from Paper 1, that considered a stationary champagne flow. The ionized gas accelerates into the \HII\ region and, subsequently, drags the dust along with it. The work in Paper 1 describes the theoretical framework to derive the properties of the dust grains introduced into the \HII\ region and provides an analysis of the IDW surrounding $\sigma$ Ori AB. In this section, we extend this work by considering the emission from morphology A as well (observed at 8 $\mu$m,  12 $\mu$m, and longwards of 160 $\mu$m). Furthermore, we present how one can use dust waves to constrain geometry (porosity/fluffiness) and the radiation pressure efficiency of the grains.

\subsection{Modeling the flow of ionized gas}\label{sec:model}

\citet{henney_2005} investigated the evaporation of a cloud illuminated by a point source, and the structure of the ionized flow emanating of the cloud surface. In their hydrodynamical models, the photo-evaporation flows are found to be steady during a large part of their evolution, in which the flow properties change only on timescales much longer than the dynamic time for flow away from the IF. In the steady phase, the flow resembles a 'champagne flow' \citep{tenorio_tagle_1979} or a 'globule flow' \citep{bertoldi_1989}, depending on the curvature of the IF. 

An IF that engulfs a flat, homogenous cloud illuminated by a point source will become concave, and the photo-evaporation flow emanating from this geometry will resemble a champagne flow structure, as the streamlines do not diverge effectively \citep{henney_2005}. In practice, the ionized gas from a concave IF will accelerate more slowly compared to, e.g., the photo-evaporation flow from a convex-shaped globule \citep{bertoldi_1989}. The relative slow acceleration of the gas will increase the effective flow thickness, $h_\m{eff}$, defined as $n_0^2$$h_\m{eff}$ = $\int$ $n^2$ d$r$, where $n_\m{0}$ is the ionized gas density and $r$ is the distance {\em from} the star {\em to} the IF. Thus, the shape of the IF (convex, flat, or concave) and the incident ionizing flux {\em uniquely} determine the structure of the resulting photo-evaporation flow. 

We envision the approach of $\sigma$ Ori to the L1630 molecular cloud as follows. We consider the L1630 molecular cloud as a flat IF, ionized by an (approaching) point source. By combining the density prescription that reproduces the observed emission from the IC 434 region (Eq. \ref{eq:densgrad}) with the results of the numerical models of \citet{henney_2005}, we developed a flow model that improves the model described in Paper 1 substantially. We restrict the modeling of the flow to the symmetry axis connecting the star and the cloud, because the flow properties are dominated by the immediate surroundings of the symmetry axis \citep{henney_2005}, as will be addressed in Sec. \ref{sec:flowparameters}. Moreover, the photo-evaporation models from \citet{henney_2005} show that in the case of a flat IF illuminated by a single point source, the streamlines do not diverge at the distance between star and cloud, so we can consider the gas flow to be moving one-dimensionally (i.e., orthogonal to the surface of the molecular cloud) constant along the line-of-sight. Thus, the symmetry axis provides a decent approximation for the (two-dimensional) structure of the flow between the cloud and the star. In short, the analysis results in an increased line-of-sight extent of the cloud compared to Paper 1, but is consisted with the numerical models of \citet{henney_2005} and comparable to the size of the L1630 molecular cloud.

The photo-evaporation flow from the L1630 molecular cloud resembles a champagne flow following the prescription of \citet{tenorio_tagle_1979}. In this scenario, the density contrast between a cloud and the inter cloud medium drives a shock into the inter cloud medium, while a rarefaction wave travels back into the cloud, setting up a champagne flow that accelerates the gas to supersonic velocities as the gas moves away from the IF. The density distribution of the champagne flow can be described by an exponential density gradient \citep{bedijn_1981,tielens_2005}:

\begin{equation}
\label{eq:densgrad}
n(r) = n_\m{0}\exp\left[\frac{r-R_\m{s}}{R_\m{0}}\right].
\end{equation} 

\noindent Here, $n_\m{0}$ is evaluated at the IF and equals $J$/$c_\m{s}$, where $J$ is the ionizing flux reaching the cloud and $c_\m{s}$ is the isothermal sound speed in the ionized gas. The value $R_\m{s}$ is the distance between the cloud and the star, and $R_\m{0}$ = $c_\m{s}$$t$ is the scale length of the flow, where $t$ is the time since the ionized gas started to stream into the \HII\ region. We follow the conclusions of \citet{henney_2005} by adopting a constant flow thickness as the star approaches the cloud, i.e., the scale height of the flow, $R_\m{0}$ (see Eq. \ref{eq:densgrad}), will adjust itself to  the distance between the IF and the star, $R_\m{s}$. In particular, the effective flow thickness for a flat IF is $h_\m{eff}$ $\approx$ 0.35$R_\m{s}$ \citep[Fig. 7]{henney_2005}.

Thus, assuming that the density gradient can be described by Eq. \ref{eq:densgrad} (Paper 1) and an effective flow thickness of $h_\m{eff}$ = 0.35$R_\m{s}$, one can solve for the scale height $R_\m{0}$ of the exponential flow by writing $n_\m{0}^2$ {\bf $h_\m{eff}$ = $\int_{0}^{R_\m{s}}$ $(n(r))^2$ $\m{d}r$}. To determine the complete structure of the flow as the ionizing star draws closer, we now only have to solve for the ionizing flux, $J$, reaching the cloud surface at any given time. The ionizing flux will increase as the star advances towards the cloud and will accelerate the IF, increasing the ionized gas density streaming into the \HII\ region. Consequently, the number of recombinations between the star and the cloud will increase and, in its turn, lower the amount of photons reaching the cloud surface. The actual amount of photons reaching the cloud is determined by integrating the number of recombinations in a column $r$ extending from the star ($r$ = 0) to the IF ($r$ = $R_\m{s}$):

\begin{equation}
\label{eq:ionflux}
J  = J_\m{0} - \int_0^{R_\m{s}} \beta_\m{B} (n(r))^2 \m{d}r = \frac{\beta
_\m{B}J^2R_\m{0}}{2c_\m{s}^2}(1-\exp\left[-2R_\m{s}/R_\m{0}\right]),
\end{equation}

\noindent where $J_\m{0}$ = $Q_\m{0}$/4$\pi$$r^2$ is the ionizing flux from the star, in the case that none of the photons are absorbed within the \HII\ region. Equation \ref{eq:ionflux} yields the solution: 

\begin{equation}
\label{eq:ionflux2}
\frac{2J_\m{0}}{J} = 1 + \sqrt{1+ \frac{Q_\m{0}\beta_\m{B}R_\m{0}}{4\pi r^2 2c_\m{s}^2}(1-\exp\left[-2R_\m{s}/R_\m{0}\right])}.
\end{equation} 

\noindent This result is similar to the one reported in \citet{bedijn_1981}, except their solution is written in terms of number density; we opted to follow the method from \citet{spitzer_1978} and provide the solution in term of stellar fluxes. With the density prescription, we can solve for the scale height $R_\m{0}$ of the flow using the previous mentioned relationship, $n_\m{0}^2$ {\bf $h_\m{eff}$ = $\int_{0}^{R_\m{s}}$ $(n(r))^2$ $\m{d}r$}. With $n_\m{0}$ = $J$/$c_\m{s}$, and $h_\m{eff}$ = 0.35$R_\m{s}$, it is easily shown that $R_\m{0}$ = 0.46$R_\m{s}$. 

\subsection{Flow parameters}\label{sec:flowparameters}

Consider the L1630 molecular cloud as a uniformly distributed, extended cloud. We take for the ionizing flux of $\sigma$ Ori AB log($Q_\m{0}$) = 47.56 photons s$^{-1}$, appropriate for an O9.5V star \citep{martins_2005}. We let the star approach the cloud at a velocity of 10 km s$^{-1}$ (Sec. \ref{sec:spaceapproach}), which after 3 Myr (the adopted age of $\sigma$ Ori; \citealt{caballero_2008}) results in the configuration seen to date: a projected distance of 3.7 pc between the L1630 cloud and $\sigma$ Ori AB. At $t$ = 0, we assume that a steady photo-evaporation flow already fills the medium between the star and cloud (see Fig. \ref{fig:flowstructure}).

\begin{figure}
\centering
\includegraphics[width=9cm]{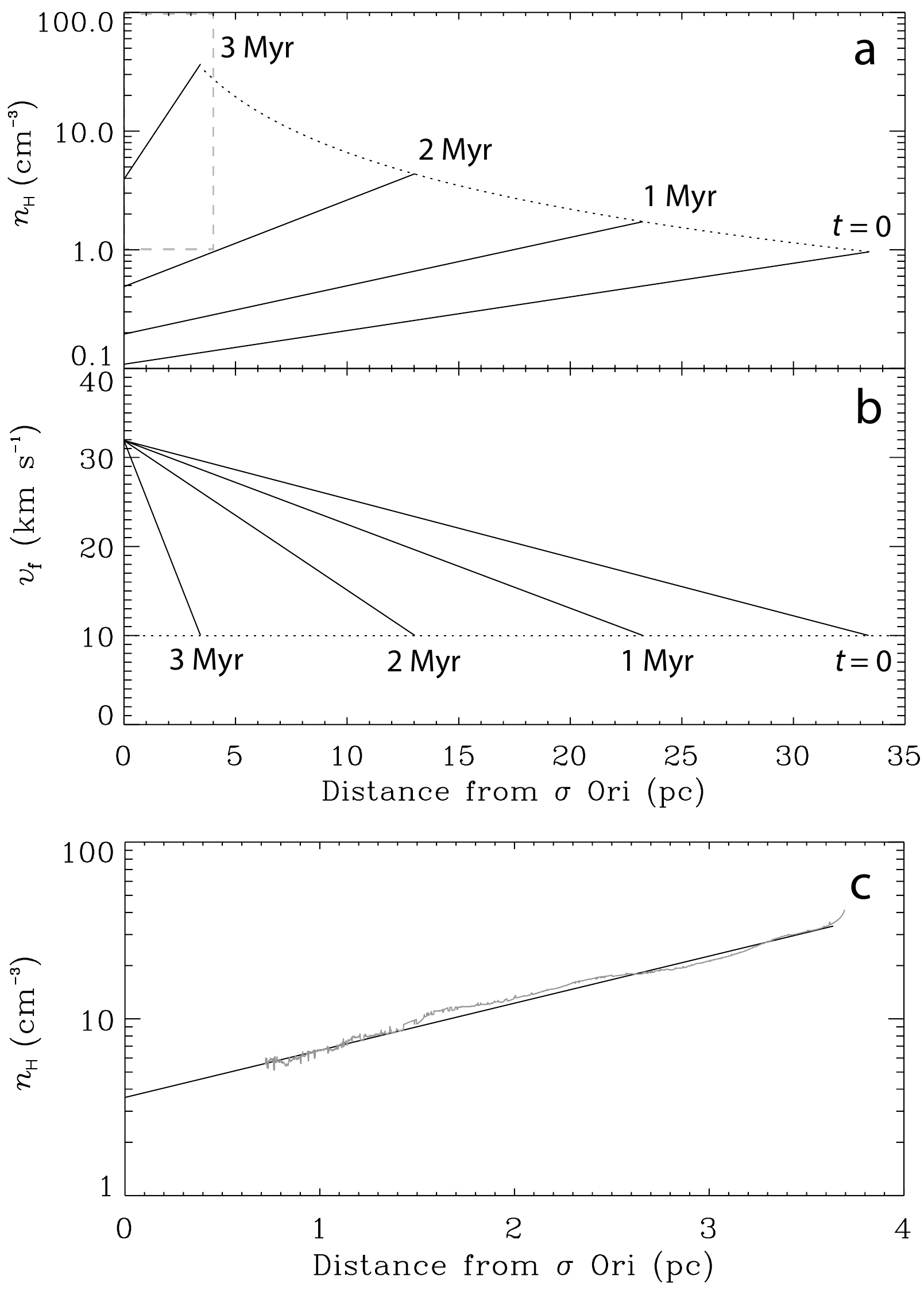} 
\caption{{\bf (a)} Density structure $n$ of the champagne flow as $\sigma$ Ori moves toward the cloud (Sec. \ref{sec:spaceapproach}). The different curves represent the density structure after time $t$: 0 Myr, 1 Myr, 2Myr, and 3 Myr. The dotted line shows the evolution of the density at the IF, $n_\m{0}$. The grey dashed box shows the outline of the region of panel c. {\bf (b)} Velocity structure of the flow. To transform this to the frame between star and cloud, one has to add the relative velocity between the star and cloud, $v_\star$ = 10 km s$^{-1}$. {\bf (c)} The density profile at $t$ = 3 Myr, overplotted with the observed density profile from the H$\alpha$ emission measure, assuming a constant scale size of the emitting gas along the line of sight of $l$ = 9 pc (see text).}
\label{fig:flowparameters}
\end{figure}

Figure \ref{fig:flowparameters} reveals the evolution of the flow parameters, $v_\m{f}$ and $n_\m{H}$, as $\sigma$ Ori travels towards the cloud at a velocity of $v_\star$ = 10 km s$^{-1}$ (Sec. \ref{sec:spaceapproach}). The velocity of the flow, $v_\m{f}$, is obtained after integration of Eq. \ref{eq:densgrad}, resulting in a linear velocity law. The initial velocity of the gas is $v_\m{f}$ = 10 km s$^{-1}$ (a D-critical front; \citealt{spitzer_1978}) and, because of the constant scale height $R_\m{0}$ \citep{henney_2005}, reaches the same maximum velocity of $v_\m{f}$ = 32 km s$^{-1}$ at all times when the gas reaches the location of the star. Taking into account the relative velocity between star and cloud, this amounts to a maximum relative velocity of 42 km s$^{-1}$. 

The density structure of the flow at $t$ = 3 Myr reaches $n_\m{H}$ = 35 cm$^{-3}$ at the cloud surface. This density is offset by a factor of $\sim$ 3 with the earlier derivation described in Paper 1. This difference follows from the choice of $Q_\m{0}$: in Paper 1, the O9.5V model atmosphere of \citet{schaerer_1997}, with an ionizing luminosity of log($Q_\m{0}$) = 48.25 photons s$^{-1}$, was taken to calculate the incident flux on the L1630 cloud. Here, we use instead the updated model atmospheres from \citet{martins_2005} and take log($Q_\m{0}$) = 47.56 photons s$^{-1}$. The decrease of ionizing photons effectively lowers $n_\m{H}$. In addition, we calculate the balance between recombinations and photo-ionization appropriate for a champagne flow structure with Eq. \ref{eq:ionflux2}, whereas in Paper 1 we had followed the work from \citet{abergel_2002} and \citet{spitzer_1978} that calculate $J_\m{0}$ in a spherical geometry. While a spherical, globule-like geometry is appropriate for the small evaporation zone for the Horsehead considered in \citet{abergel_2002}, it clearly is not applicable to the entire champagne-flow from the L1630 molecular cloud (e.g., Fig. \ref{fig:flowparameters}). The S[III] lines that were used in \citet{compiegne_2007} (and that we had adopted in Paper 1) are close to or below the critical density, and its usage to derive reliable densities is questionable. Therefore, lacking other reliable density tracers that can pin down the initial density at the IF, calculating the amount of photons that reach the molecular cloud in time, using the appropriate density laws from Eqs. \ref{eq:ionflux} and  \ref{eq:ionflux2}, is the most reliable way in constraining the flow parameters in the IC 434 nebula. In addition, we compare with numerical models from \citet{henney_2005} to obtain a better description of the ionized flow of the L1630 molecular cloud.

The lower density $n_\m{H}$ remains compatible with the observed H$\alpha$ emission measure from the IC 434 region, as the electron density depends on the scale size of the emitting gas through $n_\m{e}$ = $\sqrt{EM/l}$. Here, EM and $l$ are the emission measure and the scale size along the line-of-sight of the emitting volume, and $n_\m{e}$ (= $n_\m{H}$) is the electron density of the (fully) ionized gas. Thus, by lowering $n_\m{e}$ by a factor of 3, $l$ has to increase factor of 3$^2$ along the line-of-sight to reproduce the EM of the ionized flow, which amounts to $l$ = 9 pc at the L1630 molecular cloud, similar to the angular extent of the IC 434 region (Fig. \ref{fig:ic434}). 

While strictly speaking, Eqs. \ref{eq:ionflux} and \ref{eq:ionflux2} hold for the symmetry axis only (extending from the star to the cloud radially), the bulk of the emission measure originates from the region within one unit distance $R_\m{s}$ from the symmetry axis \citep[Fig. 5a]{henney_2005} and, therefore, the dust emission from the \HII\ region will largely be located within this region. In the case of IC 434, this amounts to an emitting area of $\sim$ 7.5 pc centered on the symmetry axis, consistent with our adopted line-of-sight extent of the L1630 molecular cloud. Even though our adopted geometry and flow structure is consistent with numerical results from \citet{henney_2005} and the size of the L1630 cloud on the sky, we note that the observed EM can be used with {\em any} $l$ to match an exponential density gradient inside the \HII\ region, depending on the choice of $R_\m{0}$. We look into this dependency by investigating our results after varying the scale height $R_\m{0}$ in Sec. \ref{sec:discussion}.

In summary, we have updated the photo-evaporation model compared to the one described in Paper 1 by using (1) a more recent model atmosphere, (2) a better prescription of calculating the 'insulating' layer of recombining material streaming of the molecular cloud (Eqs. \ref{eq:ionflux} \& \ref{eq:ionflux2}), (3) a comparison with numerical results for photo-evaporation flows from \citet{henney_2005}, and (4) and allowed for the time evolution of the flow as $\sigma$ Ori approaches the cloud. The exponential density gradient (Fig. \ref{fig:flowparameters}a) and the adopted scale height (Sec. \ref{sec:model}) reproduces the observed H$\alpha$ emission measure well if the IC 434 region measures 9 pc along the line of sight, in agreement with the photo-evaporation models of \citet{henney_2005}. Note, that while the updated flow structure does affect some of the results derived in Paper 1, the main conclusions do not change and the derived grain properties are only affected marginally, as we will discuss in Sec. \ref{sec:results}.

\subsection{The formation of dust waves}\label{sec:dustwave}

In this section, we briefly review the dust wave phenomenon. Following Paper 1, the equation of motion for dust is written as:

\begin{equation}
\label{eq:eqmotiondust}
m_\m{d}v_\m{d}\frac{dv_\m{d}}{dr} = -\frac{\sigma_\m{d} \bar{Q}_\m{rp} L_{\star}}{4 \pi cr^2} + 2\sigma_\m{d}kTn_i\frac{8s}{3 \sqrt{\pi} } \left(1 + \frac{9\pi}{64}s_i^2\right)^{1/2}.
\end{equation}

\noindent The first term on the right hand side is the radiation pressure force, $F_\m{rad}$, and contains: the geometrical cross-section of the grain, $\sigma_\m{d}$; the radiation pressure efficiency (averaged over the incident radiation field), $\bar{Q}_\m{rp}$; and the speed of light, $c$. The second term is the collisional drag force  \citep{draine_1979}, $F_\m{drag}$, and contains: the Boltzmann constant, $k$; the temperature and number density of the gas (of species $i$),  $T$ and $n_\m{i}$; and $s_i = (m_i v_\m{drift}^2/2kT)^{1/2}$, with $v_\m{drift}$ the drift velocity of the dust through the gas. Furthermore, $m_\m{d}$ is the dust mass and $v_\m{d}$ is the dust velocity.

As the grains flow along with the gas into the \HII\ region through the drag force, $F_\m{drag}$, the radiation pressure of the star, $F_\m{rad}$, will act on a dust grain contained within the flow and will cause it to lose momentum. As the dust is slowed down, it will be pushed through the gas with drift velocity $v_\m{drift}$, transferring momentum towards the gas. We consider collisional drag only, as Coulomb interactions do not seem to dominate the drag term for dust-gas momentum transfer inside IC 434 (Paper 1). Eventually, for a dust grain approaching the star radially (i.e., zero impact parameter), the dust will be stopped at the equilibrium radius, $r_\m{eq}$, upstream with respect to the star, where $|F_\mathrm{rad}$/$F_\mathrm{drag}|$ equals unity. If at this point the amount of momentum transfer between gas and dust is insufficient for the gas to be dynamically perturbed, gas and dust will decouple significantly, resulting in an increase of dust upstream relative to the star as the gas flows along unhindered: the formation of a {\em dust wave}. Dust grains approaching the star with a non-zero impact parameter will be pushed around the star, resulting in an arc-shaped structure. For a detailed description of the physics of a dusty photo-evaporation flow we refer to (Paper 1).

\begin{figure}
\centering
\includegraphics[width=9cm]{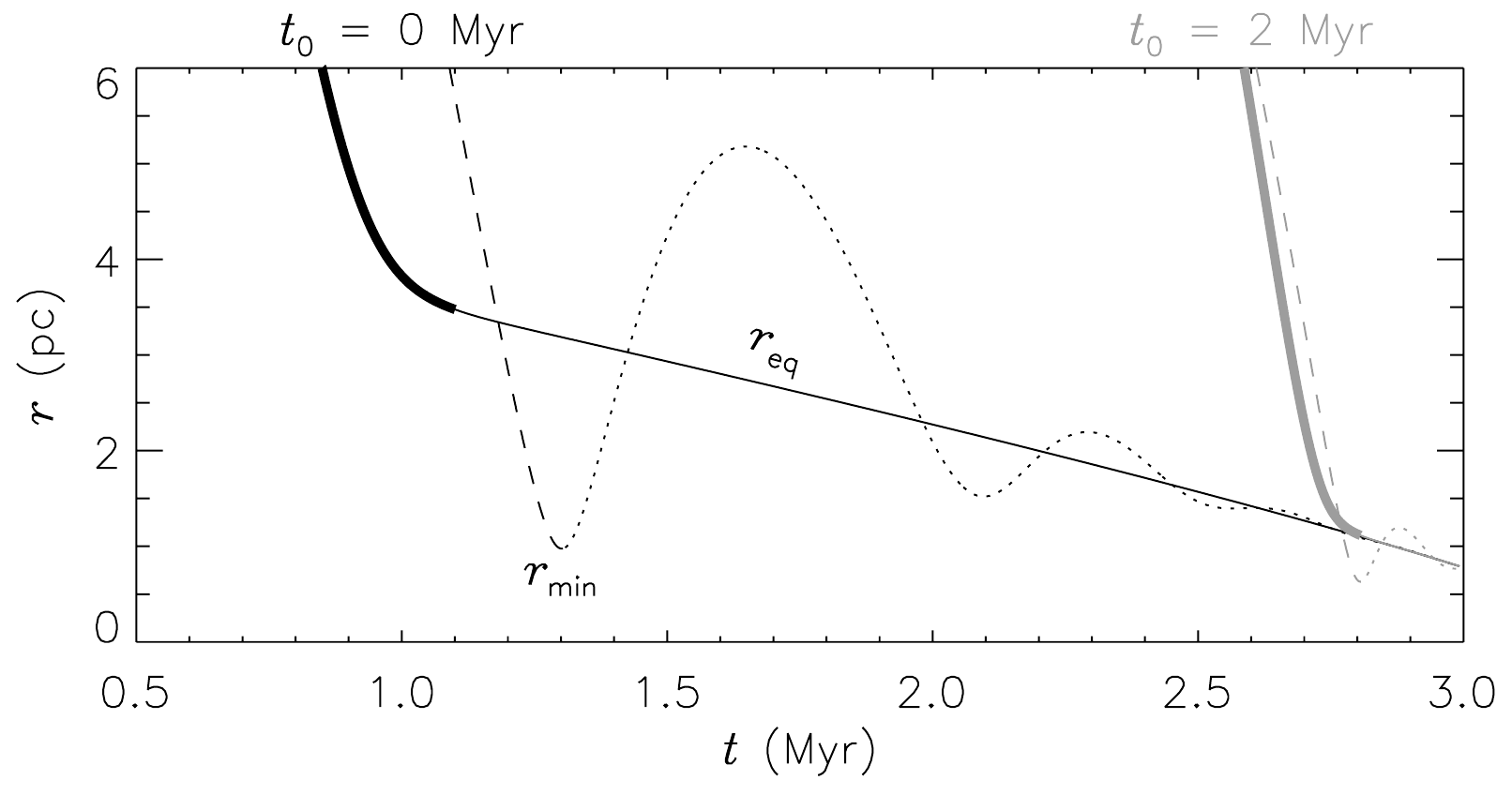} 
\caption{The trajectories of two grains with different value of geometrical cross-section over dust mass (see text), $\xi$, exemplifying the difference between the equilibrium radius, $r_\m{eq}$ (solid line; high $\xi$), and the minimum radius, $r_\m{min}$ (dashed/dotted line; low $\xi$) versus time $t$ for dust approaching the star radially. Both curves assume $\bar{Q}_\m{rp}$ = 2. Black curves represent dust entering the \HII\ region at $t_\m{0}$ = Myr, while grey curves correspond to the same situation at $t_\m{0}$ = 2 Myr.}
\label{fig:req}
\end{figure}

\subsection{The equilibrium- and minimum radius of a dust wave}\label{sec:reqvsrmin}

The radiation pressure opacity, $\kappa_\m{rp}$, determines the trajectory of a dust grain as it moves toward and passes the star :

\begin{equation}
\label{eq:radopacity}
\kappa_\m{rp} = \bar{Q}_\m{rp} \sigma_\m{d} / m_\m{d} \equiv \bar{Q}_\m{rp} \xi.
\end{equation}

\noindent In Paper 1, it was shown that the radiation pressure force of $\sigma$ Ori AB stratifies grains ahead of the star in a {\em bounded} region at 0.1 pc $\textless$ $d$ $\textless$ 0.4 pc for a given range of $\kappa_\m{rp}$. The analysis was limited to grains with $\bar{Q}_\m{rp}$ = 1, from which a size distribution of the grains contained in the dust wave was inferred assuming compact particles, i.e., $m_\m{d}$ = 4/3$\pi$$a^3$$\rho_\m{s}$, with $m_\m{d}$ dust mass, $a$ the grain radius and $\rho_\m{s}$ the specific density. However, given Eq. \ref{eq:radopacity}, we can obtain the same values of $\kappa_\m{rp}$ by allowing for variation of $Q_\m{rp}$, and a corresponding change in $\xi$. For example, a grain with $\bar{Q}_\m{rp}$ = 1 and a compact geometry will have exactly the same value of $\kappa_\m{rp}$ as a grain with $\bar{Q}_\m{rp}$ = 0.1 and $\xi$ increased by a factor of 10 (a fluffy/porous geometry). We attempt to break this degeneracy by comparing our modeling results with observations (Sec. \ref{sec:results}).

The inner boundary of a dust wave is determined by grains that have a low ratio of $\xi$. Considering grains at same $m_\m{d}$, smaller values of $\xi$ correspond to more compact grains. Due to inertia, these grains will slowly accelerate and, moreover, tend to overshoot $r_\m{eq}$ due to their large intrinsic momentum, reaching the minimum radius $r_\m{min}$ constrained by $\xi$ (Fig. \ref{fig:req}). 

For sufficient high $\xi$, the grains wil be {\em momentum coupled} with the gas of the flow. In this case, a negligible amount of momentum is used to accelerate the grains, from which follows $F_\m{rad}$ = $F_\m{drag}$ (Eq. \ref{eq:eqmotiondust}). Then, all momentum absorbed by the grains will be passed to the gas, and the grains will have a steady drift velocity through the gas {\em independent} of grain size \citep{kwok_1975,tielens_1983}. The equilibrium radius, $r_\m{eq}$, is populated by grains that are momentum coupled, whose exact value will only depend on the flow parameters, $v_\m{f}$ and $n$, and the value of the radiation pressure opacity $\bar{Q}_\m{rp}$. High values of $\xi$ either corresponds to very small dust grains (e.g., VSGs), or large grains with a low specific density, such as porous grains or fluffy aggregates. It is good to note here that porosity and fluffiness are not directly related. Whereas the porosity is a measure of the volume filling factor of a grain \citep[e.g.,][]{bazell_1990} and contains no information on its morphological structure, a fluffy or filamentary grain is often described by a fractal dimension \citep[e.g.,][]{fogel_1998}. Grains with the same porosity can have very different fractal dimension. However in this work, only $\xi$ enters Eq. \ref{eq:eqmotiondust}, and we can not distinguish between porosity and fluffiness. 

Figure \ref{fig:req} depicts the difference between $r_\m{min}$ and $r_\m{eq}$ ($\bar{Q}_\m{rp}$ = 2). The curve for $r_\m{min}$ is calculated for a 1000 \AA\ compact silicate grain (log($\xi$) = 4.03). In contrast, the curve for $r_\m{eq}$ is calculated for a grain of similar mass, but with a geometrical cross-section $\sigma_\m{d}$ increased by a factor of 100 (log($\xi$) = 6.03), to mimic a highly porous or fluffy grain that is momentum-coupled with the gas. The equilibrium radius $r_\m{eq}$ shrinks with $t$, because the density and velocity of the ionized gas increases, pushing the dust towards the star as $F_\m{drag}$ rises. We plot trajectories for dust emanating from the L1630 molecular cloud at $t_\m{0}$ = 0 Myr, at which point $F_\m{drag}$ is still relatively small (Fig. \ref{fig:flowparameters}): the compact grains overshoot significantly, and the difference $r_\m{eq}$ - $r_\m{min}$ is high. The effect of inertia is also seen in the difference in time $t$ at which the grains reach $r_\m{eq}$ and $r_\m{min}$: the compact grains are accelerated more slowly, thus reaching $r_\m{min}$ at a later stage compared to the grains that are coupled to the gas and reach $r_\m{eq}$. However, the importance of the inertia of the grains will decrease with $t$ as $\sigma$ Ori approaches the cloud and the ionized gas flow develops in strength, exemplified by the curves for dust launched at $t_\m{0}$ = 2 Myr. Note that the damped oscillation of $r_\m{min}$ around $r_\m{eq}$ is a peculiarity of the streamline approaching the star radially. In reality, dust along this streamline will flow around the star when the dust has reached $r_\m{min}$ \citep{ochsendorf_2014b} and, therefore, only the dashed part of $r_\m{min}$ is physically relevant. The same applies to momentum coupled grains: after the grains reach $r_\m{eq}$, they will be pushed past the star. The thin part of the solid line reveals $r_\m{eq}$ for all momentum-coupled grains that arrive at $r_\m{eq}$ between 1.1 Myr $\leq$ $t$ $\leq$ 2.8 Myr.

In summary, for given flow parameters and adopting a dust size distribution, dust grains will be stratified in a region ahead of the star, stretching from $r_\m{eq}$ (small or porous grains) towards $r_\m{min}$ (compact, large grains), depending on $\xi$. These radii mark the location where the dust reaches minimal velocities with respect to the star, resulting in high dust densities: a dust wave. The radiation pressure efficiency, $\bar{Q}_\m{rp}$, effectively offsets $r_\m{eq}$ and $r_\m{min}$ with respect to the star, as $\bar{Q}_\m{rp}$ only enters $F_\m{rad}$ in Eq. \ref{eq:eqmotiondust}. Values for $\bar{Q}_\m{rp}$ depend on grain properties and the incident radiation field \citep{li_2001,draine_2011}. For a radiation field typical of that of an O9.5V star, compact, spherical grains of size $\sim$ 1000 \AA\, can have an efficiency factor as high as $\bar{Q}_\m{rp}$ $\approx$ 2. In contrast, small grains of 10 \AA\ have $\bar{Q}_\m{rp}$ $\approx$ 0.1, while for the largest grain sizes with 2$\pi$$a$ $\gg$ $\lambda$, $\bar{Q}_\m{rp}$ approaches unity.

\section{Modeling results}\label{sec:results}

For the flow parameters estimated in Sec. \ref{sec:flowparameters}, and $\bar{Q}_\m{rp}$ = 1.5, which is a reasonable estimate for classical silicate and graphite grains larger than $\sim$ 0.01 $\mu$m \citep[see][Fig. 7]{draine_2011}, at $t$ = 3 Myr (the present situation) the equilibrium radius equals $r_\m{eq}$ = 0.7 pc. At the same time, $r_\m{min}$ = 0.3 pc for 0.25 $\mu$m grains, typically the largest grains available in the size distribution of the diffuse interstellar medium \citep{mathis_1977,weingartner_2001}. Thus, for $\bar{Q}_\m{rp}$ = 1.5 and assuming an ISM size distribution, dust will be stratified between 0.3 pc $\textless$ $r$ $\textless$ 0.7 pc ahead of the star, which is to be compared to the observed projected distances of the dust waves. Surely, the actual difference between $\sigma$ Ori and the dust waves could be larger than the projected distance due to an inclination of the system. However, this effect should be small, as observations indicate that $\sigma$ Ori AB and the Horsehead (and, therefore, the L1630 molecular cloud associated with the Horsehead) lie approximately at the same distance \citep{habart_2005}. 

In contrast to the region bounded by $r_\m{min}$ and $r_\m{eq}$ as described above, the IR emission from the IC 434 region (Fig. \ref{fig:flowstructure}) revealed two clearly separated morphologies with corresponding dust arcs at a projected distance of $d$ = 1 pc (the ODW; morphology A) and $d$ = 0.1 pc (the IDW; morphology B). While in our model, $r_\m{min}$ can be lowered by introducing large grains ($a$ $\textgreater$ 0.25 $\mu$m) into the size distribution (these have lower $r_\m{min}$ due to their large momentum), $r_\m{eq}$ determines the outer region of the dust wave, which can only be increased to match the observed value (1 pc) by allowing for a larger value of $\bar{Q}_\m{rp}$ (Eq. \ref{eq:radopacity}). However by increasing $\bar{Q}_\m{rp}$, consequently, $r_\m{min}$ will rise and divert from 0.1 pc. Therefore, a continuous interstellar size distribution will not reproduce the two-folded nature of morphology A versus morphology B, as described in Sec. \ref{sec:observations}, and the segregation of the dust can only be reproduced by invoking a bimodal dust distribution, separating the dust into their current morphologies and observational characteristics.

\subsection{A bimodal dust distribution}\label{sec:results_bimodal}

\begin{figure}
\centering
\includegraphics[width=9cm]{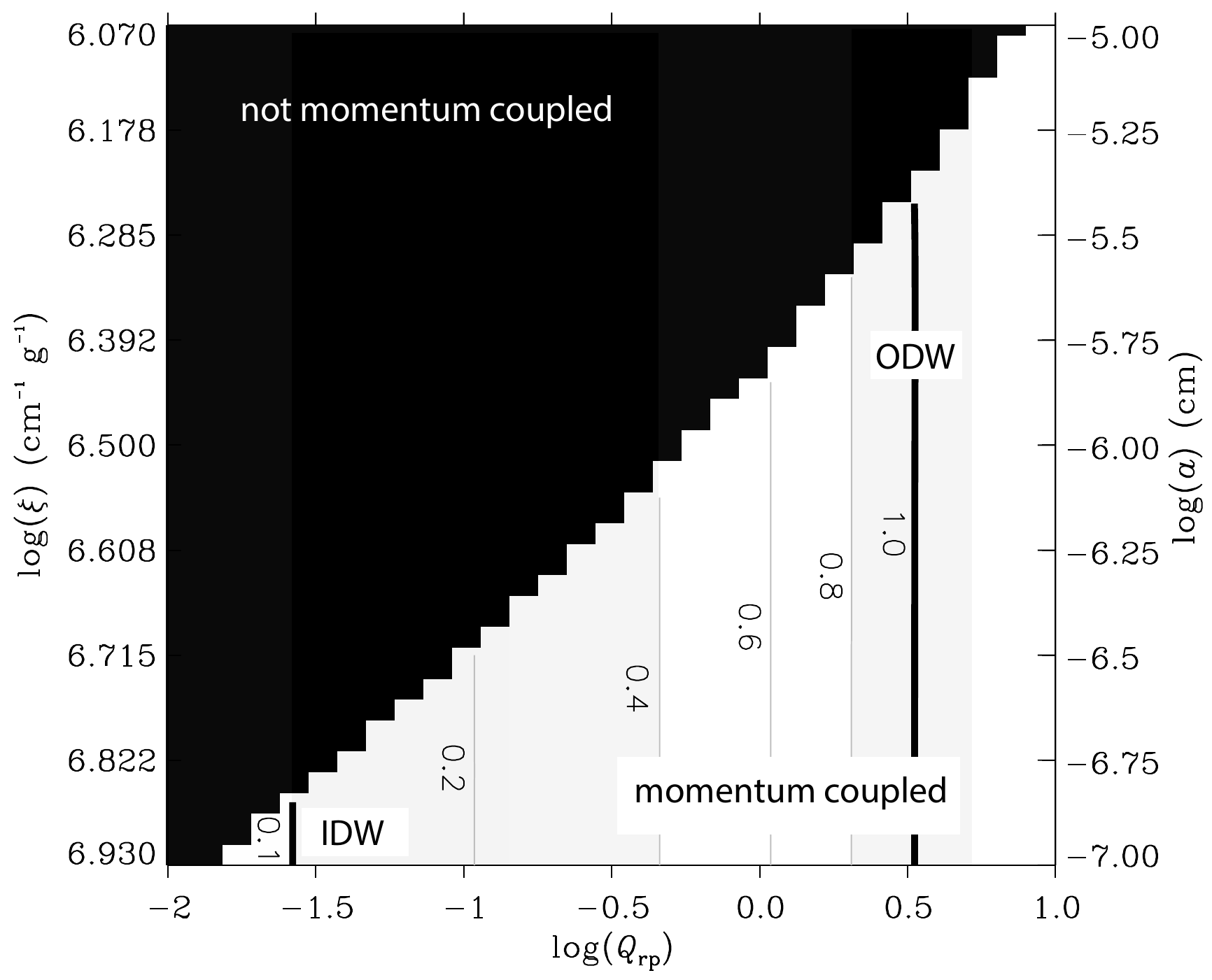} 
\caption{Location of the equilibrium radius $r_\m{eq}$ in pc (contours) of momentum coupled grains at $t$ = 3 Myr (present situation), evaluated for a grid of radiation pressure efficiencies $\bar{Q}_\m{rp}$ and ratios of geometrical cross-section to the mass of the grain, $\xi$. The thick black contours represent $r_\m{eq}$ = 0.1 and  $r_\m{eq}$ = 1.0, respectively, corresponding to the peak intensities of the inner dust wave (IDW) and outer dust wave (ODW). The light-shaded areas mark the boundaries in projected distance at which the dust waves are observed Fig. (\ref{fig:flowstructure}). The black area represent grains that overshoot $r_\m{eq}$ to reach $r_\m{min}$, and are thus not momentum coupled. The vertical axis on the right denotes the grain radius if the grain mass is distributed in a compact geometry.}
\label{fig:qvsratio}
\end{figure}

Section \ref{sec:reqvsrmin} discussed that momentum-coupled grains reach $r_\m{eq}$, the outer radius of a dust wave, and that $\bar{Q}_\m{rp}$ effectively offsets the interaction region between $r_\m{eq}$ and $r_\m{min}$ from the star (Eq. \ref{eq:radopacity}). In Fig \ref{fig:qvsratio}, we visualize this, and plot $r_\m{eq}$ at $t$ = 3 Myr for a grid of $\bar{Q}_\m{rp}$ and $\xi$. This diagram reveals the necessary grain properties to place momentum-coupled grains at the observed positions of the ODW and the IDW, respectively. Above a certain value of $\xi$, the grains will obtain sufficient inertia such that they are no longer momentum-coupled. The division between both regimes shifts to large values of $\xi$, as $\bar{Q}_\m{rp}$ increases. This is because radiation pressure becomes more efficient, prohibiting the grains from overshooting $r_\m{eq}$.

\subsubsection{Dust population A}\label{sec:morphologya}

The ODW at $d$ = 1 pc places a tight constraint on the nature of grains responsible for morphology A. We have demonstrated that for the flow parameters in IC 434 (Sec. \ref{sec:flowparameters}) and $\bar{Q}_\m{rp}$ =1.5, $r_\m{eq}$ = 0.7 pc. Therefore, we conclude that in order to place dust at the ODW, the grains have to be momentum coupled, for these grains account for the outer regions of the dust wave measured from the star. Subsequently, we can search for the value $\bar{Q}_\m{rp}$ that places $r_\m{eq}$ at $d$ = 1 pc in Fig. \ref{fig:qvsratio}: momentum coupled grains with radiation pressure efficiency 2 $\leq$ $\bar{Q}_\m{rp}$ $\leq$ 5 could reproduce the ODW around 1 pc. 

The required $\bar{Q}_\m{rp}$ is larger than commonly accepted \citep{li_2001}. Indeed, the above described constraints (momentum-coupled, high $Q_\m{rp}$), combined with the results from Sec. \ref{sec:colddust} (low $T$, insensitive to $G_\m{0}$), seem to be incompatible with any grain material used in the dust models of, e.g., \citet{draine_2007, compiegne_2011}. Here, we note that the grains need to be large to account for the cold temperatures (Sec. \ref{sec:colddust}), yet in order to assure momentum coupling with the gas, log($\xi$) $\gtrsim$ 6.3 (Fig. \ref{fig:qvsratio}). This implies that a large silicate grain with a mass equal to a compact grain with $a$ = 1 $\mu$m and $\rho_\m{s}$ = 3.5 g cm$^{-3}$ will need an increase of geometrical cross-section by a factor of $\sim$ 50 to assure momentum-coupling with the gas. At $a$ = 0.1 $\mu$m, this factor decreases to 17. We defer a further discussion on the remarkable properties characterizing morphology A to Sec. \ref{sec:discussion}.

\subsubsection{Dust population B}\label{sec:morphologyb}

The presence of the IDW at 0.1 $\textless$ $d$ $\textless$ 0.4 pc, exhibiting the observational characteristics described in Sec. \ref{sec:obsproperties}, indicates the presence of dust grains different in nature compared to morphology A. Here, there are two scenarios to consider. 

First, the Spitzer/IRS LL spectra, shown in Fig. \ref{fig:llregions}, revealed that the 24 $\mu$m emission from the \HII\ region outside of the IDW is dominated by carbonaceous VSGs (Sec. \ref{sec:vsgs}). Within this framework, the IDW can be formed by these small, compact grains, that are easily momentum-coupled to the gas, yet are less efficient in absorbing the radiation pressure from the star. Figure \ref{fig:qvsratio} shows that momentum-coupled grains with radiation pressure efficiency 0.03 $\leq$ $\bar{Q}_\m{rp}$ $\leq$ 0.50 could reproduce the IDW at 0.1 $\textless$ $d$ $\textless$ 0.4 pc. These values for $\bar{Q}_\m{rp}$ are consistent with values obtained for classical compact grains of size $a$ $\textless$ 0.01 $\mu$m \citep{li_2001,draine_2011}.

Second, large, compact BGs that overshoot $r_\m{eq}$ due to inertia can explain the presence of the IDW (this scenario was already described in Paper 1). For large, compact grains, $\bar{Q}_\m{rp}$ approaches unity \citep{draine_2007,draine_2011}. Then, for the flow parameters adopted, grains of size $a$ = 0.5 $\mu$m reach the inner parts of the dust wave at $d$ $\sim$ 0.1 pc (in Paper 1, we derived $a$ = 0.33 $\mu$m due to the different flow parameters; Sec. \ref{sec:flowparameters}). This is a very large size above the typical cut-off for grains in the ISM ($a$ $\approx$ 0.25 $\mu$m; \citet{weingartner_2001}). The exact dust grain size to reach the inner part of the wave depends on the adopted flow parameters, as well as the momentum of the grain. 

The grains that populate the IDW can therefore be explained by either VSGs with low $\bar{Q}_\m{rp}$ that have reached $r_\m{eq}$, or large, compact BGs for which $\bar{Q}_\m{rp}$ reaches unity. The reduced UV-to-NIR opacity (Sec. \ref{sec:opacity}) and the observed dust color temperature of the IDW at 0.1 pc (73 K - 82 K, Sec. \ref{sec:obsproperties}) hints towards large grains (Tab. \ref{tab:dusttemp}), consistent with the conclusion from Paper 1: a carbonaceous VSG of size 1 $\times$ 10$^{-3}$ $\mu$m will reach $T_\m{eq}$ =  170 K at a projected distance of 0.1 pc towards the star, while a large grain of 0.5 $\mu$m reaches a temperature $T_\m{eq}$ = 75 K.

\section{Discussion}\label{sec:discussion}

The simultaneous presence of two clearly separated dust morphologies inside IC 434, with separate dust waves at projected distance of 0.1 pc and 1.0 pc, can not be explained with a continuous size distribution like that observed for selected regions of the diffuse interstellar medium. Combining modeling (Sec. \ref{sec:results}) with the constrains derived through observations (Sec. \ref{sec:obsproperties}), we have argued that the dust in IC 434 is bimodal in nature.

\subsection{Derived dust properties}

Population A is characterized observationally by a cold temperature (20 K - 26 K; depending on the spectral emissivity $\beta$) that is below the equilibrium temperature of classical dust grains predicted by DUSTEM \citep{compiegne_2011}. Moreover, the temperature of this population as derived from a MBB fit to the SED only increases by $\sim$ 15\% (Sec. \ref{sec:colddust}), while the incident stellar flux increases by an order of magnitude: DUSTEM predicts a temperature increase of 50\% for typical ISM grains (Sec. \ref{sec:obsproperties}). We find a discrepancy between UV and IR inferred mass of dust contained in the ODW (Sec. \ref{sec:observations}). Modeling of the photo-evaporation flow reveals that the corresponding dust wave at 1.0 pc can only be reproduced if the grains are momentum-coupled with the gas, which can be obtained by increasing the geometrical cross-section of the grains by factors of 17 and 50 for grains with a mass equivalent to a compact (silicate) grain of 0.1 $\mu$m and 1.0 $\mu$m, respectively. Furthermore, they need to be highly effective in absorbing radiation pressure ($\bar{Q}_\m{rp}$ $\geq$ 2; Sec. \ref{sec:morphologya}). In contrast, population B is characterized observationally by mid-IR at $\sim$ 24 emission throughout the \HII\ region, complemented with the IDW that peaks in intensity at 0.1 pc. The emission in the \HII\ region outside of the IDW is caused by VSGs out-of-equilibrium with the radiation field, supported by the IRS spectra. However, the dust wave at 0.1 pc can be explained both by VSGs, but also by BGs that are invisible throughout the \HII\ region and only start to emit at mid-IR wavelengths close to the star as the grains heat up (consistent with the IRS spectrum of region 7). The modeling in Sec. \ref{sec:morphologyb} reveals that either momentum-coupled grains with $\bar{Q}_\m{rp}$ $\textless$ 1, or very large compact grains with $\bar{Q}_\m{rp}$ = 1 can cause the dust wave at 0.1 pc. The color temperature of 73 K - 82 K, and the reduced UV-to-NIR opacity argue for the latter scenario (consisted with Paper 1). 

Population B is consistent with grain properties used in current dust models \citep[e.g.,][]{li_2001, compiegne_2011}, and we connect this population to the warm dust observed in many \HII\ regions \citep{chini_1986,paladini_2012, ochsendorf_2014b}. However, we note that Paper 1 argued that the charging of these grains deviate significantly from that predicted by theory \citep[e.g.,][]{draine_1979}, consistent with the dust waves in the RCW 120 and RCW 82 \HII\ regions \citep{ochsendorf_2014b}. However, the characteristics of population A are perhaps even more remarkable, and a critical review seems appropriate to place our findings in context with recent developments in our understanding of dust modeling and evolution. 

The cold temperature of morphology A (Sec. \ref{sec:colddust}) and the reduced UV-to-NIR opacity (Sec. \ref{sec:opacity}) argue for large grains, and the requirement for momentum-coupling with the gas indicates a fluffy or porous structure (Sec. \ref{sec:results_bimodal}). Indeed, detailed modeling  \citep{ossenkopf_1994, ormel_2009} and observational evidence  \citep{steinacker_2010} indicate that grains reach large sizes inside dense cores ($\textgreater$ 1 $\mu$m), from which the presence of large aggregates seems intuitive. The observations may reflect their past evolution in the cloud as the grains are evaporated into the \HII\ region (see for a similar conclusion \citealt{martin_2012}). However, the constant temperature of population A in a varying radiation field, derived from the MBB fits, will pose a problem to every available dust model. A possible explanation is that the size distribution is heavily affected inside the \HII\ region through processing by the intense radiation field of $\sigma$ Ori AB. In this scenario, the radiation field destroys the smallest grains, which suppresses the small end of the size distribution. This effect was already seen in Sec. \ref{sec:obsproperties}, where a population of VSGs did not seem to survive the transition from the well-shielded PDR of the Horsehead nebula into the \HII\ region. An alternative explanation stems from the results from \citet{abergel_2013} and \citet{aniano_2014}, who show that the peak wavelength of the SED does not trace the intensity of the radiation field only (Sec. \ref{sec:colddust}): the authors argue that this may be a signature of grain properties that are evolving as the grains transit from one region to another. This argumentation could in principle be applied to the transition of the dust in IC 434 between the dense molecular cloud phase to the \HII\ region. Still, the derived (high) $\bar{Q}_\m{rp}$ for this population remains unexplained: it is questionable if dust aggregates are indeed efficient radiation absorbers. For example, \citet{bazell_1990} conclude that the optical properties of fractal grains do not change significantly compared to their compact counterparts, which would imply $\bar{Q}_\m{rp}$ $\leq$ 2 \citep{li_2001,draine_2011}. The detailed optical properties of grains containing chemical inhomogeneities and differing fractal dimensions remain poorly understood; in this respect, more theoretical research that constrain the optical properties of dust aggregates is necessary.

In conclusion, while the analysis of the dust budget in the IC 434 region provides insight into the properties of the grains residing in the ionized gas, some of the results are incompatible with current modeling and should be subject to further questioning. Indeed, evidence is accumulating that the 'standard model' of interstellar dust is incapable to explain the variations of grain properties when moving from one region of the ISM to another (for a recent review, see \citealt{jones_2014}). The findings of this work support this view, and indicates that our knowledge of dust may be far from complete. Perhaps, we are tracing unknown grain materials. Further laboratory experiments on cosmic dust analogues and theoretical calculations on the optical properties of large aggregates must be used in designing the next generations of dust models, which should also incorporate the evolution of grain materials and properties during the lifecycle of the grains in the ISM.

\subsection{Limitations of the model}

The derived dust parameters depend on the adopted stellar parameters, as well as the model of the approach of $\sigma$ Ori towards the L1630 molecular cloud. Luckily, the density structure of the flow is well constrained by observations and reveals an exponential dependency with distance from the cloud. However, our choice for the ionizing flux of $\sigma$ Ori AB, log($Q_\m{0}$) = 47.56 photons s$^{-1}$ \citep{martins_2005}, is low compared to values predicted by other model atmospheres. For example, the model atmosphere from \citet{schaerer_1997} has log($Q_\m{0}$) = 48.25 photons s$^{-1}$ for a O9.5V spectral type star. Because we consider a constant effective flow thickness \citep{henney_2005}, a higher value of $Q_\m{0}$ would only increase the density of the ionized gas, thereby increasing $F_\m{drag}$. An increase of ionizing flux would therefore lead to an increase of our derived values for $\bar{Q}_\m{rp}$, and vice versa. 

\begin{figure}
\centering
\includegraphics[width=9cm]{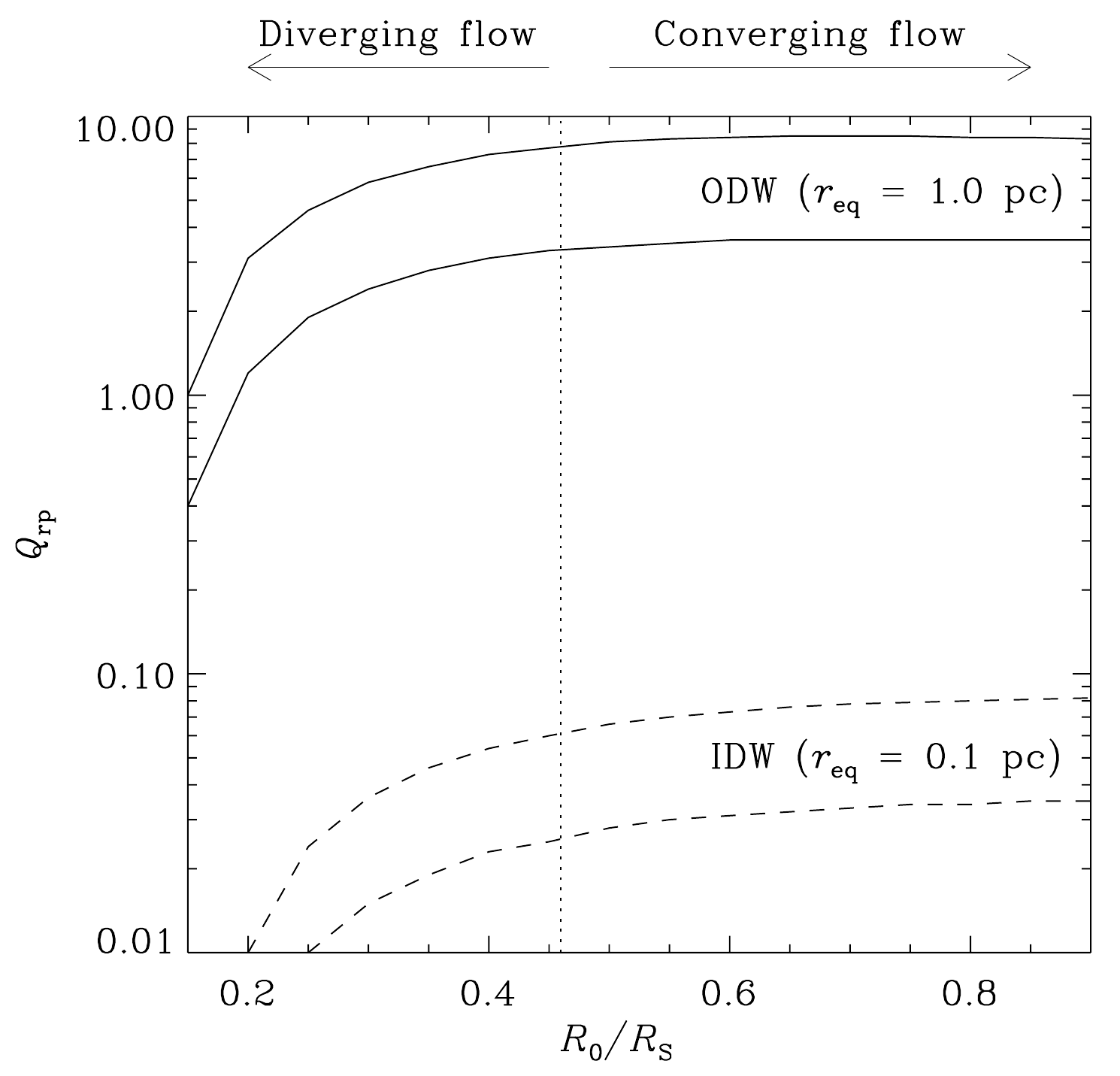} 
\caption{The influence of the adopted structure for the champagne flow on the derived values of the radiation pressure efficiency, $\bar{Q}_\m{rp}$. Plotted is $\bar{Q}_\m{rp}$ versus the scale height of the flow, $R_\m{0}$, normalized to the distance star - cloud, $R_\m{s}$. The dashed line marks the scale height, corresponding to a flat ionization front, and is the one assumed in our calculations. Smaller values of $R_\m{0}$ correspond to increasingly diverging flows, characterized by a fast acceleration of the gas. Larger values of $R_\m{0}$ correspond to thick flows that accelerate more slowly. Plotted is $\bar{Q}_\m{rp}$ such that the equilibrium radius equals $r_\m{eq}$ = 1.0 (solid lines) and  $r_\m{eq}$ = 0.1 (dashed lines). The upper curves correspond to an ionizing flux of log($Q_\m{0}$) = 48.25 photons s$^{-1}$ \citep{schaerer_1997}, the lower curves correspond to an ionizing flux of log($Q_\m{0}$) = 47.56 photons s$^{-1}$ \citep{martins_2005}.}
\label{fig:flowthick}
\end{figure}

The dependency on the flow scale height $R_\m{0}$ is less straightforward, and we opted to solve this numerically. As discussed in Sec. \ref{sec:flowparameters}, the adopted value of $R_\m{0}$ (= 0.46$R_\m{s}$) is appropriate for a photo-evaporation flow emanating from a flat IF \citep{henney_2005}. This choice of $R_\m{0}$ reproduces the H$\alpha$ emission measure with constant scale size $l$. In Fig \ref{fig:flowthick}, we vary $R_\m{0}$ and plot the dependency on our results for $\bar{Q}_\m{rp}$. Only when $R_\m{0}$ is smaller than 0.46$R_\m{s}$, corresponding to a diverging flow, are our results for $\bar{Q}_\m{rp}$ significantly affected. This is because in this case, the gas reaches higher velocities, yet has a lower density. The {\em total} amount of momentum transfer between the dust and gas in time decreases, lowering $F_\m{drag}$ and, therefore, lowering $\bar{Q}_\m{rp}$ to place $r_\m{eq}$ at the required distances. This result was already observed in \citet{ochsendorf_2014b}: momentum transfer between gas and dust is most efficient in slow-moving, high-density photo-evaporation flows. In this respect, at high $R_\m{0}$, the flow parameters do not change significantly and barely affects $\bar{Q}_\m{rp}$ anymore. 

To conclude, the observed H$\alpha$ emission measure is well described by an exponential power law, $n$ $\propto$ $\exp\left[-r/R_\m{0}\right]$, which implies that that the gas does not accelerate efficiently like that in a globule flow, where $n$ $\propto$ $r^{-2}$ \citep{bertoldi_1989}. Therefore, low values for $R_\m{0}$ ($\textless$ 0.46$R_\m{s}$) are unlikely for the photo-evaporation flow in IC 434, while our results for $\bar{Q}_\m{rp}$ will not be significantly altered for $R_\m{0}$ $\textgreater$ 0.46$R_\m{s}$.

\subsection{PAH emission from the ionized gas}\label{sec:pahs}

In Sec. \ref{sec:dustdistribution}, we have shown that morphology A is both traced by thermal emission of large dust grains at $\lambda$ $\geq$ 160 $\mu$m, and by IR fluorescence of PAHs, emitting at 8 $\mu$m - 12 $\mu$m. In the framework of dusty photo-evaporation flows, it seems peculiar that PAHs and large dust grains are spatially correlated. PAHs are easily dragged along by a flow of ionized gas, whereas large grains will lag behind due to inertia. We propose that PAH molecules are formed {\em inside} the \HII\ region through shattering upon grain-grain collisions between the separate grain populations in the flow, that couple different to the gas and move at non-identical velocities. Sputtering is not efficient for the physical conditions inside the IC 434 region (Paper 1). Once being formed, the PAHs are destroyed quickly through photo-dissociation after UV photon absorption, expected at the intensity of the radiation field \citep{tielens_2005}. This could explain why PAHs and large grains are seen co-spatial, and results in the absence of PAHs within $\sim$ 1 pc from the star. This cavity is likely seen in the evolution of the PAH spectrum (Fig. \ref{fig:pahevolution}), where a discontinuity is observed in the 7.7 $\mu$m over 11.3 $\mu$m intensity ratio between region 6 and region 7. A detailed model describing the collision- and fragmentation process, and the evolution of the PAH emission spectrum inside the \HII\ region is beyond the scope of this paper, and will be presented in a forthcoming work (Ochsendorf et al., in prep.).

\section{Conclusions}\label{sec:conclusions}

We have carried out a combined observational- and modeling study of the IC 434 region and the GS206-17+13 shell that surrounds it. We characterized the dust emission from the \HII\ region (Secs. \ref{sec:observations} \& \ref{sec:results}), with the use of an updated flow model (Sec. \ref{sec:introducingdust}) that resulted from an analysis of the large-scale structure of the Belt region seen to date (Sec. \ref{sec:1bshell}). The important findings are listed below:

\begin{enumerate}
\item[1.] The GS206-17+13 shell is driven by the Orion OB1b association, rather than the $\sigma$ Ori cluster, derived through a study of space motions, and by comparing the kinetic energy of the expanding shell to the mechanical output of the Orion OB1b members and the weak-wind system of $\sigma$ Ori AB (Sec. \ref{sec:1bshell}). The space motion of $\sigma$ Ori AB is directed toward the L1630 molecular cloud, and can be traced back to the Orion OB1c association, in line with the suggestion of \citet{bally_2008} that assigned the $\sigma$ Ori cluster to the younger Orion OB1c association. This would solve the apparent mismatch between the age of $\sigma$ Ori ($\sim$ 3 Myr; \citealt{caballero_2008}) to the Orion OB1b association ($\textgreater$ 5 Myr; \citealt{bally_2008}). Besides that the aforementioned analysis of the Belt region results in a flow model necessary to derive the grain properties in the \HII\ region, it provides new insight in the history of the Belt region that will be of great interest for research on the Orion OB association and stellar feedback in OB associations. The Orion region is far from fully understood, and with the upcoming results from the Gaia mission, its study provides a benchmark case that has implications for the evolution of OB associations in general.

\item[3.] The global structure of the IR emission revealed two separate morphologies (morphology A \& B) associated with a dust wave at 1.0 pc and 0.1 pc, respectively. We have attributed this segregation to a bimodal dust distribution (population A \& B), and have characterized both populations through their observational properties and through modeling of the grain trajectories.

\item[4.] A dust wave stratifies dust grains in a region stretching from $r_\m{eq}$ (momentum-coupled grains) towards $r_\m{min}$ (non-momentum-coupled grains), depending on the geometrical cross-section over dust mass, $\xi$. The radiation pressure efficiency, $\bar{Q}_\m{rp}$, effectively offsets $r_\m{eq}$ and $r_\m{min}$ from the star. 

\item[5.] Population A has a cold dust temperature (20 K - 27 K; depending on the spectral emissivity) derived from MBB fits, below the equilibrium temperature of 'classical' grains at the incident radiation field (20 K - 55 K for grains 0.1 $\mu$m - 1.0 $\mu$m) predicted by DUSTEM. PAH emission is seen co-spatial with population A, and the increasing charge state of the molecules reveal that both the PAHs and population A reside within the ionized gas. The dust wave at 1 pc, associated with population A, can only be reproduced if the grains are momentum-coupled with the gas (log($\xi$) $\textgreater$ 6.25), for which the grain need to have a significant increased geometrical cross section compared to compact grains. We have argued that the porous or fluffy aggregates that are formed inside the L1630 molecular cloud are good candidates to explain the above described characteristics of population A. However, the high radiation pressure efficiency ($\bar{Q}_\m{rp}$ $\geq$ 2), and the evolution of the temperature that does not seem to trace an increase of radiation field only, can not be explained by current dust models, but confirms recent results by, e.g., {\em Planck} \citep{abergel_2013,aniano_2014}, indicating that our understanding of interstellar dust may be limited \citep{jones_2014}.

\item[6.] Population B is reminiscent of warm dust, often observed towards the centers of \HII\ regions \citep{chini_1986,paladini_2012,ochsendorf_2014b}. Here, we reveal that it is characterized by mid-IR emission around 24 $\mu$m from VSGs throughout the \HII\ region (region 1 - 6), and a dust wave at 0.1 pc from the star where the dust reaches TE with $T_\m{d}$ $\sim$ 73 K - 82 K. The dust wave at 0.1 pc can be reproduced if the grains are either momentum-coupled small grains (log($\xi$) $\textgreater$ 6.9, $\bar{Q}_\m{rp}$ $\textless$ 1), or large, compact grains ($a$ $\approx$ 0.5 $\mu$m, $\bar{Q}_\m{rp}$ = 1) that approach the star closely due to inertia. These values for $\bar{Q}_\m{rp}$ are consistent with prediction from theory. The reduced UV-to-NIR opacity, and the observed dust color temperature of the IDW are consistent with the BG scenario, confirming the results from Paper 1.

\item[7.] The presence of PAH emission inside the \HII\ region is linked to shattering of grain material upon grain-grain collisions.  

\end{enumerate}

\begin{acknowledgements} 
Studies of interstellar dust and chemistry at Leiden Observatory are supported through advanced ERC grant 246976 from the European Research Council, through a grant by the Dutch Science Agency, NWO, as part of the Dutch Astrochemistry Network, and through the Spinoza premie from the Dutch Science Agency, NWO.
\end{acknowledgements}

\bibliographystyle{aa} 
\bibliography{bimodal.bib} 

\begin{thebibliography}{91}
\expandafter\ifx\csname natexlab\endcsname\relax\def\natexlab#1{#1}\fi

\bibitem[{{Abergel} {et~al.}(2002){Abergel}, {Bernard}, {Boulanger},
  {Cesarsky}, {Falgarone}, {Jones}, {Miville-Deschenes}, {Perault}, {Puget},
  {Huldtgren}, {Kaas}, {Nordh}, {Olofsson}, {Andr{\'e}}, {Bontemps}, {Casali},
  {Cesarsky}, {Copet}, {Davies}, {Montmerle}, {Persi}, \&
  {Sibille}}]{abergel_2002}
{Abergel}, A., {Bernard}, J.~P., {Boulanger}, F., {et~al.} 2002, \aap, 389, 239

\bibitem[{{Allamandola} {et~al.}(1999){Allamandola}, {Hudgins}, \&
  {Sandford}}]{allamandola_1999}
{Allamandola}, L.~J., {Hudgins}, D.~M., \& {Sandford}, S.~A. 1999, \apjl, 511,
  L115

\bibitem[{{Anderson} {et~al.}(2012){Anderson}, {Zavagno}, {Deharveng},
  {Abergel}, {Motte}, {Andr{\'e}}, {Bernard}, {Bontemps}, {Hennemann}, {Hill},
  {Rod{\'o}n}, {Roussel}, \& {Russeil}}]{anderson_2012}
{Anderson}, L.~D., {Zavagno}, A., {Deharveng}, L., {et~al.} 2012, \aap, 542,
  A10

\bibitem[{{Andr{\'e}} {et~al.}(2010){Andr{\'e}}, {Men'shchikov}, {Bontemps},
  {K{\"o}nyves}, {Motte}, {Schneider}, {Didelon}, {Minier}, {Saraceno},
  {Ward-Thompson}, {di Francesco}, {White}, {Molinari}, {Testi}, {Abergel},
  {Griffin}, {Henning}, {Royer}, {Mer{\'{\i}}n}, {Vavrek}, {Attard},
  {Arzoumanian}, {Wilson}, {Ade}, {Aussel}, {Baluteau}, {Benedettini},
  {Bernard}, {Blommaert}, {Cambr{\'e}sy}, {Cox}, {di Giorgio}, {Hargrave},
  {Hennemann}, {Huang}, {Kirk}, {Krause}, {Launhardt}, {Leeks}, {Le Pennec},
  {Li}, {Martin}, {Maury}, {Olofsson}, {Omont}, {Peretto}, {Pezzuto}, {Prusti},
  {Roussel}, {Russeil}, {Sauvage}, {Sibthorpe}, {Sicilia-Aguilar}, {Spinoglio},
  {Waelkens}, {Woodcraft}, \& {Zavagno}}]{andre_2010}
{Andr{\'e}}, P., {Men'shchikov}, A., {Bontemps}, S., {et~al.} 2010, \aap, 518,
  L102

\bibitem[{{Aniano} {et~al.}(2011){Aniano}, {Draine}, {Gordon}, \&
  {Sandstrom}}]{aniano_2011}
{Aniano}, G., {Draine}, B.~T., {Gordon}, K.~D., \& {Sandstrom}, K. 2011, \pasp,
  123, 1218

\bibitem[{{Bally}(2008)}]{bally_2008}
{Bally}, J. 2008, {Overview of the Orion Complex}, ed. B.~{Reipurth}, 459

\bibitem[{{Bauschlicher}(2002)}]{bauschlicher_2002}
{Bauschlicher}, Jr., C.~W. 2002, \apj, 564, 782

\bibitem[{{Bazell} \& {Dwek}(1990)}]{bazell_1990}
{Bazell}, D. \& {Dwek}, E. 1990, \apj, 360, 142

\bibitem[{{Beckwith} {et~al.}(1990){Beckwith}, {Sargent}, {Chini}, \&
  {Guesten}}]{beckwith_1990}
{Beckwith}, S.~V.~W., {Sargent}, A.~I., {Chini}, R.~S., \& {Guesten}, R. 1990,
  \aj, 99, 924

\bibitem[{{Bedijn} \& {Tenorio-Tagle}(1981)}]{bedijn_1981}
{Bedijn}, P.~J. \& {Tenorio-Tagle}, G. 1981, \aap, 98, 85

\bibitem[{{Bernard} {et~al.}(2010){Bernard}, {Paradis}, {Marshall}, {Montier},
  {Lagache}, {Paladini}, {Veneziani}, {Brunt}, {Mottram}, {Martin},
  {Ristorcelli}, {Noriega-Crespo}, {Compi{\`e}gne}, {Flagey}, {Anderson},
  {Popescu}, {Tuffs}, {Reach}, {White}, {Benedetti}, {Calzoletti}, {Digiorgio},
  {Faustini}, {Juvela}, {Joblin}, {Joncas}, {Mivilles-Deschenes}, {Olmi},
  {Traficante}, {Piacentini}, {Zavagno}, \& {Molinari}}]{bernard_2010}
{Bernard}, J.-P., {Paradis}, D., {Marshall}, D.~J., {et~al.} 2010, \aap, 518,
  L88

\bibitem[{{Bertoldi}(1989)}]{bertoldi_1989}
{Bertoldi}, F. 1989, \apj, 346, 735

\bibitem[{{Blaauw}(1964)}]{blaauw_1964}
{Blaauw}, A. 1964, \araa, 2, 213

\bibitem[{{Blomme} {et~al.}(2002){Blomme}, {Prinja}, {Runacres}, \&
  {Colley}}]{blomme_2002}
{Blomme}, R., {Prinja}, R.~K., {Runacres}, M.~C., \& {Colley}, S. 2002, \aap,
  382, 921

\bibitem[{{Brown} {et~al.}(1994){Brown}, {de Geus}, \& {de Zeeuw}}]{brown_1994}
{Brown}, A.~G.~A., {de Geus}, E.~J., \& {de Zeeuw}, P.~T. 1994, \aap, 289, 101

\bibitem[{{Brown} {et~al.}(1995){Brown}, {Hartmann}, \& {Burton}}]{brown_1995}
{Brown}, A.~G.~A., {Hartmann}, D., \& {Burton}, W.~B. 1995, \aap, 300, 903

\bibitem[{{Caballero}(2007{\natexlab{a}})}]{caballero_2007b}
{Caballero}, J.~A. 2007{\natexlab{a}}, Astronomische Nachrichten, 328, 917

\bibitem[{{Caballero}(2007{\natexlab{b}})}]{caballero_2007}
{Caballero}, J.~A. 2007{\natexlab{b}}, \aap, 466, 917

\bibitem[{{Caballero}(2008)}]{caballero_2008}
{Caballero}, J.~A. 2008, \mnras, 383, 750

\bibitem[{{Cardelli} {et~al.}(1989){Cardelli}, {Clayton}, \&
  {Mathis}}]{cardelli_1989}
{Cardelli}, J.~A., {Clayton}, G.~C., \& {Mathis}, J.~S. 1989, \apj, 345, 245

\bibitem[{{Cesarsky} {et~al.}(2000){Cesarsky}, {Lequeux}, {Ryter}, \&
  {G{\'e}rin}}]{cesarsky_2000}
{Cesarsky}, D., {Lequeux}, J., {Ryter}, C., \& {G{\'e}rin}, M. 2000, \aap, 354,
  L87

\bibitem[{{Chevalier}(1974)}]{chevalier_1974}
{Chevalier}, R.~A. 1974, \apj, 188, 501

\bibitem[{{Chini} {et~al.}(1986){Chini}, {Kruegel}, \& {Kreysa}}]{chini_1986}
{Chini}, R., {Kruegel}, E., \& {Kreysa}, E. 1986, \aap, 167, 315

\bibitem[{{Co{\c s}kuno{\v g}lu} {et~al.}(2011){Co{\c s}kuno{\v g}lu}, {Ak},
  {Bilir}, {Karaali}, {Yaz}, {Gilmore}, {Seabroke}, {Bienaym{\'e}},
  {Bland-Hawthorn}, {Campbell}, {Freeman}, {Gibson}, {Grebel}, {Munari},
  {Navarro}, {Parker}, {Siebert}, {Siviero}, {Steinmetz}, {Watson}, {Wyse}, \&
  {Zwitter}}]{coskonoglu_2011}
{Co{\c s}kuno{\v g}lu}, B., {Ak}, S., {Bilir}, S., {et~al.} 2011, \mnras, 412,
  1237

\bibitem[{{Compi{\`e}gne} {et~al.}(2007){Compi{\`e}gne}, {Abergel},
  {Verstraete}, {Reach}, {Habart}, {Smith}, {Boulanger}, \&
  {Joblin}}]{compiegne_2007}
{Compi{\`e}gne}, M., {Abergel}, A., {Verstraete}, L., {et~al.} 2007, \aap, 471,
  205

\bibitem[{{Compi{\`e}gne} {et~al.}(2011){Compi{\`e}gne}, {Verstraete}, {Jones},
  {Bernard}, {Boulanger}, {Flagey}, {Le Bourlot}, {Paradis}, \&
  {Ysard}}]{compiegne_2011}
{Compi{\`e}gne}, M., {Verstraete}, L., {Jones}, A., {et~al.} 2011, \aap, 525,
  A103

\bibitem[{{Cox} {et~al.}(2012){Cox}, {Kerschbaum}, {van Marle}, {Decin},
  {Ladjal}, {Mayer}, {Groenewegen}, {van Eck}, {Royer}, {Ottensamer}, {Ueta},
  {Jorissen}, {Mecina}, {Meliani}, {Luntzer}, {Blommaert}, {Posch},
  {Vandenbussche}, \& {Waelkens}}]{cox_2012}
{Cox}, N.~L.~J., {Kerschbaum}, F., {van Marle}, A.-J., {et~al.} 2012, \aap,
  537, A35

\bibitem[{{de Bruijne}(2012)}]{debruijne_2012}
{de Bruijne}, J.~H.~J. 2012, \apss, 341, 31

\bibitem[{{de Zeeuw} {et~al.}(1999){de Zeeuw}, {Hoogerwerf}, {de Bruijne},
  {Brown}, \& {Blaauw}}]{de_zeeuw_1999}
{de Zeeuw}, P.~T., {Hoogerwerf}, R., {de Bruijne}, J.~H.~J., {Brown}, A.~G.~A.,
  \& {Blaauw}, A. 1999, \aj, 117, 354

\bibitem[{{Draine}(2011)}]{draine_2011}
{Draine}, B.~T. 2011, \apj, 732, 100

\bibitem[{{Draine} \& {Li}(2007)}]{draine_2007}
{Draine}, B.~T. \& {Li}, A. 2007, \apj, 657, 810

\bibitem[{{Draine} \& {Salpeter}(1979)}]{draine_1979}
{Draine}, B.~T. \& {Salpeter}, E.~E. 1979, \apj, 231, 77

\bibitem[{{Ehlerov{\'a}} \& {Palou{\v s}}(2005)}]{ehlerova_2005}
{Ehlerov{\'a}}, S. \& {Palou{\v s}}, J. 2005, \aap, 437, 101

\bibitem[{{Everett} \& {Churchwell}(2010)}]{everett_2010}
{Everett}, J.~E. \& {Churchwell}, E. 2010, \apj, 713, 592

\bibitem[{{Flagey} {et~al.}(2011){Flagey}, {Boulanger}, {Noriega-Crespo},
  {Paladini}, {Montmerle}, {Carey}, {Gagn{\'e}}, \& {Shenoy}}]{flagey_2011}
{Flagey}, N., {Boulanger}, F., {Noriega-Crespo}, A., {et~al.} 2011, \aap, 531,
  A51

\bibitem[{{Fogel} \& {Leung}(1998)}]{fogel_1998}
{Fogel}, M.~E. \& {Leung}, C.~M. 1998, \apj, 501, 175

\bibitem[{{Galliano} {et~al.}(2008){Galliano}, {Madden}, {Tielens}, {Peeters},
  \& {Jones}}]{galliano_2008}
{Galliano}, F., {Madden}, S.~C., {Tielens}, A.~G.~G.~M., {Peeters}, E., \&
  {Jones}, A.~P. 2008, \apj, 679, 310

\bibitem[{{Gibb} {et~al.}(1995){Gibb}, {Little}, {Heaton}, \&
  {Lehtinen}}]{gibb_1995}
{Gibb}, A.~G., {Little}, L.~T., {Heaton}, B.~D., \& {Lehtinen}, K.~K. 1995,
  \mnras, 277, 341

\bibitem[{{Habart} {et~al.}(2005){Habart}, {Abergel}, {Walmsley}, {Teyssier},
  \& {Pety}}]{habart_2005}
{Habart}, E., {Abergel}, A., {Walmsley}, C.~M., {Teyssier}, D., \& {Pety}, J.
  2005, \aap, 437, 177

\bibitem[{{Habing}(1968)}]{habing_1968}
{Habing}, H.~J. 1968, \bain, 19, 421

\bibitem[{{Henney} {et~al.}(2005){Henney}, {Arthur}, \&
  {Garc{\'{\i}}a-D{\'{\i}}az}}]{henney_2005}
{Henney}, W.~J., {Arthur}, S.~J., \& {Garc{\'{\i}}a-D{\'{\i}}az}, M.~T. 2005,
  \apj, 627, 813

\bibitem[{{Hildebrand}(1983)}]{hildebrand_1983}
{Hildebrand}, R.~H. 1983, \qjras, 24, 267

\bibitem[{{Hony} {et~al.}(2001){Hony}, {Van Kerckhoven}, {Peeters}, {Tielens},
  {Hudgins}, \& {Allamandola}}]{hony_2001}
{Hony}, S., {Van Kerckhoven}, C., {Peeters}, E., {et~al.} 2001, \aap, 370, 1030

\bibitem[{{Inoue}(2002)}]{inoue_2002}
{Inoue}, A.~K. 2002, \apj, 570, 688

\bibitem[{{Jones}(2014)}]{jones_2014}
{Jones}, A. 2014, ArXiv e-print 1411.6666

\bibitem[{{Juvela} \& {Ysard}(2012)}]{juvela_2012}
{Juvela}, M. \& {Ysard}, N. 2012, \aap, 539, A71

\bibitem[{{Kudritzki} {et~al.}(1999){Kudritzki}, {Puls}, {Lennon}, {Venn},
  {Reetz}, {Najarro}, {McCarthy}, \& {Herrero}}]{kudritzki_1999}
{Kudritzki}, R.~P., {Puls}, J., {Lennon}, D.~J., {et~al.} 1999, \aap, 350, 970

\bibitem[{{Kwok}(1975)}]{kwok_1975}
{Kwok}, S. 1975, \apj, 198, 583

\bibitem[{{Li} \& {Draine}(2001)}]{li_2001}
{Li}, A. \& {Draine}, B.~T. 2001, \apj, 554, 778

\bibitem[{{Li} \& {Draine}(2002)}]{li_2002}
{Li}, A. \& {Draine}, B.~T. 2002, \apj, 572, 232

\bibitem[{{Maddalena} {et~al.}(1986){Maddalena}, {Morris}, {Moscowitz}, \&
  {Thaddeus}}]{maddalena_1986}
{Maddalena}, R.~J., {Morris}, M., {Moscowitz}, J., \& {Thaddeus}, P. 1986,
  \apj, 303, 375

\bibitem[{{Martin} {et~al.}(2012){Martin}, {Roy}, {Bontemps},
  {Miville-Desch{\^e}nes}, {Ade}, {Bock}, {Chapin}, {Devlin}, {Dicker},
  {Griffin}, {Gundersen}, {Halpern}, {Hargrave}, {Hughes}, {Klein}, {Marsden},
  {Mauskopf}, {Netterfield}, {Olmi}, {Patanchon}, {Rex}, {Scott}, {Semisch},
  {Truch}, {Tucker}, {Tucker}, {Viero}, \& {Wiebe}}]{martin_2012}
{Martin}, P.~G., {Roy}, A., {Bontemps}, S., {et~al.} 2012, \apj, 751, 28

\bibitem[{{Martins} {et~al.}(2005){Martins}, {Schaerer}, \&
  {Hillier}}]{martins_2005}
{Martins}, F., {Schaerer}, D., \& {Hillier}, D.~J. 2005, \aap, 436, 1049

\bibitem[{{Mathis} {et~al.}(1977){Mathis}, {Rumpl}, \&
  {Nordsieck}}]{mathis_1977}
{Mathis}, J.~S., {Rumpl}, W., \& {Nordsieck}, K.~H. 1977, \apj, 217, 425

\bibitem[{{Mennella} {et~al.}(1995){Mennella}, {Colangeli}, \&
  {Bussoletti}}]{mennella_1995}
{Mennella}, V., {Colangeli}, L., \& {Bussoletti}, E. 1995, \aap, 295, 165

\bibitem[{{Najarro} {et~al.}(2011){Najarro}, {Hanson}, \&
  {Puls}}]{najarro_2011}
{Najarro}, F., {Hanson}, M.~M., \& {Puls}, J. 2011, \aap, 535, A32

\bibitem[{{Ochsendorf} {et~al.}(2014{\natexlab{a}}){Ochsendorf}, {Cox},
  {Krijt}, {Salgado}, {Bern{\'e}}, {Bernard}, {Kaper}, \&
  {Tielens}}]{ochsendorf_2014a}
{Ochsendorf}, B.~B., {Cox}, N.~L.~J., {Krijt}, S., {et~al.} 2014{\natexlab{a}},
  \aap, 563, A65

\bibitem[{{Ochsendorf} {et~al.}(2014{\natexlab{b}}){Ochsendorf}, {Verdolini},
  {Cox}, {Bern{\'e}}, {Kaper}, \& {Tielens}}]{ochsendorf_2014b}
{Ochsendorf}, B.~B., {Verdolini}, S., {Cox}, N.~L.~J., {et~al.}
  2014{\natexlab{b}}, \aap, 566, A75

\bibitem[{{Ormel} {et~al.}(2009){Ormel}, {Paszun}, {Dominik}, \&
  {Tielens}}]{ormel_2009}
{Ormel}, C.~W., {Paszun}, D., {Dominik}, C., \& {Tielens}, A.~G.~G.~M. 2009,
  \aap, 502, 845

\bibitem[{{Ossenkopf} \& {Henning}(1994)}]{ossenkopf_1994}
{Ossenkopf}, V. \& {Henning}, T. 1994, \aap, 291, 943

\bibitem[{{Paladini} {et~al.}(2012){Paladini}, {Umana}, {Veneziani},
  {Noriega-Crespo}, {Anderson}, {Piacentini}, {Pinheiro Gon{\c c}alves},
  {Paradis}, {Tibbs}, {Bernard}, \& {Natoli}}]{paladini_2012}
{Paladini}, R., {Umana}, G., {Veneziani}, M., {et~al.} 2012, \apj, 760, 149

\bibitem[{{Paradis} {et~al.}(2011){Paradis}, {Paladini}, {Noriega-Crespo},
  {Lagache}, {Kawamura}, {Onishi}, \& {Fukui}}]{paradis_2011}
{Paradis}, D., {Paladini}, R., {Noriega-Crespo}, A., {et~al.} 2011, \apj, 735,
  6

\bibitem[{{Perryman} {et~al.}(2001){Perryman}, {de Boer}, {Gilmore}, {H{\o}g},
  {Lattanzi}, {Lindegren}, {Luri}, {Mignard}, {Pace}, \& {de
  Zeeuw}}]{perryman_2001}
{Perryman}, M.~A.~C., {de Boer}, K.~S., {Gilmore}, G., {et~al.} 2001, \aap,
  369, 339

\bibitem[{{Perryman} {et~al.}(1997){Perryman}, {Lindegren}, {Kovalevsky},
  {Hoeg}, {Bastian}, {Bernacca}, {Cr{\'e}z{\'e}}, {Donati}, {Grenon},
  {Grewing}, {van Leeuwen}, {van der Marel}, {Mignard}, {Murray}, {Le Poole},
  {Schrijver}, {Turon}, {Arenou}, {Froeschl{\'e}}, \&
  {Petersen}}]{perryman_1997}
{Perryman}, M.~A.~C., {Lindegren}, L., {Kovalevsky}, J., {et~al.} 1997, \aap,
  323, L49

\bibitem[{{Planck Collaboration} {et~al.}(2014{\natexlab{a}}){Planck
  Collaboration}, {Abergel}, {Ade}, {Aghanim}, {Alves}, {Aniano},
  {Armitage-Caplan}, {Arnaud}, {Ashdown}, {Atrio-Barandela}, \&
  et~al.}]{abergel_2013}
{Planck Collaboration}, {Abergel}, A., {Ade}, P.~A.~R., {et~al.}
  2014{\natexlab{a}}, \aap, 571, A11

\bibitem[{{Planck Collaboration} {et~al.}(2014{\natexlab{b}}){Planck
  Collaboration}, {Ade}, {Aghanim}, {Alves}, {Armitage-Caplan}, {Arnaud},
  {Ashdown}, {Atrio-Barandela}, {Aumont}, {Aussel}, \&
  et~al.}]{planck_collaboration1_2014}
{Planck Collaboration}, {Ade}, P.~A.~R., {Aghanim}, N., {et~al.}
  2014{\natexlab{b}}, \aap, 571, A1

\bibitem[{{Planck Collaboration} {et~al.}(2011){Planck Collaboration}, {Ade},
  {Aghanim}, {Arnaud}, {Ashdown}, {Aumont}, {Baccigalupi}, {Balbi}, {Banday},
  {Barreiro}, \& et~al.}]{planck_col_2011}
{Planck Collaboration}, {Ade}, P.~A.~R., {Aghanim}, N., {et~al.} 2011, \aap,
  536, A23

\bibitem[{{Plank Collaboration} {et~al.}(2014){Plank Collaboration}, {Ade},
  {Aghanim}, {Alves}, {Aniano}, {Arnaud}, {Ashdown}, {Aumont}, {Baccigalupi},
  {Banday}, {Barreiro}, {Bartolo}, {Battaner}, {Benabed}, {Benoit-Levy},
  {Bernard}, {Bersanelli}, {Bielewicz}, {Bonaldi}, {Bonavera}, {Bond},
  {Borrill}, {Bouchet}, {Boulanger}, {Burigana}, {Butler}, {Calabrese},
  {Cardoso}, {Catalano}, {Chamballu}, {Chiang}, {Christensen}, {Clements},
  {Colombi}, {Colombo}, {Couchot}, {Crill}, {Curto}, {Cuttaia}, {Danese},
  {Davies}, {Davis}, {de Bernardis}, {de Rosa}, {de Zotti}, {Delabrouille},
  {Dickinson}, {Diego}, {Dole}, {Donzelli}, {Dore}, {Douspis}, {Draine},
  {Ducout}, {Dupac}, {Efstathiou}, {Elsner}, {Ensslin}, {Eriksen}, {Falgarone},
  {Finelli}, {Forni}, {Frailis}, {Fraisse}, {Franceschi}, {Frejsel},
  {Galeotta}, {Galli}, {Ganga}, {Ghosh}, {Giard}, {Gjerlow}, {Gonzalez-Nuevo},
  {Gorski}, {Gregorio}, {Gruppuso}, {Guillet}, {Hansen}, {Hanson}, {Harrison},
  {Henrot-Versille}, {Hernandez-Monteagudo}, {Herranz}, {Hildebrandt}, {Hivon},
  {Holmes}, {Hovest}, {Huffenberger}, {Hurier}, {Jaffe}, {Jaffe}, {Jones},
  {Keihanen}, {Keskitalo}, {Kisner}, {Kneissl}, {Knoche}, {Kunz},
  {Kurki-Suonio}, {Lagache}, {Lamarre}, {Lasenby}, {Lattanzi}, {Lawrence},
  {Leonardi}, {Levrier}, {Liguori}, {Lilje}, {Linden-Vornle}, {Lopez-Caniego},
  {Lubin}, {Macias-Perez}, {Maffei}, {Maino}, {Mandolesi}, {Maris}, {Marshall},
  {Martin}, {Martinez-Gonzalez}, {Masi}, {Matarrese}, {Mazzotta}, {Melchiorri},
  {Mendes}, {Mennella}, {Migliaccio}, {Miville-Deschenes}, {Moneti}, {Montier},
  {Morgante}, {Mortlock}, {Munshi}, {Murphy}, {Naselsky}, {Natoli},
  {Norgaard-Nielsen}, {Novikov}, {Novikov}, {Oxborrow}, {Pagano}, {Pajot},
  {Paladini}, {Paoletti}, {Pasian}, {Perdereau}, {Perotto}, {Perrotta},
  {Pettorino}, {Piacentini}, {Piat}, {Plaszczynski}, {Pointecouteau},
  {Polenta}, {Ponthieu}, {Popa}, {Pratt}, {Prunet}, {Puget}, {Rachen}, {Reach},
  {Rebolo}, {Reinecke}, {Remazeilles}, {Renault}, {Ristorcelli}, {Rocha},
  {Roudier}, {Rubio-Martin}, {Rusholme}, {Sandri}, {Santos}, {Scott},
  {Spencer}, {Stolyarov}, {Sudiwala}, {Sunyaev}, {Sutton}, {Suur-Uski},
  {Sygnet}, {Tauber}, {Terenzi}, {Toffolatti}, {Tomasi}, {Tristram}, {Tucci},
  {Umana}, {Valenziano}, {Valiviita}, {Van Tent}, {Vielva}, {Villa}, {Wade},
  {Wandelt}, {Wehus}, {Ysard}, {Yvon}, {Zacchei}, \& {Zonca}}]{aniano_2014}
{Plank Collaboration}, {Ade}, P.~A.~R., {Aghanim}, N., {et~al.} 2014, ArXiv
  e-prints

\bibitem[{{Povich} {et~al.}(2007){Povich}, {Stone}, {Churchwell}, {Zweibel},
  {Wolfire}, {Babler}, {Indebetouw}, {Meade}, \& {Whitney}}]{povich_2007}
{Povich}, M.~S., {Stone}, J.~M., {Churchwell}, E., {et~al.} 2007, \apj, 660,
  346

\bibitem[{{Reynolds} \& {Ogden}(1979)}]{reynolds_1979}
{Reynolds}, R.~J. \& {Ogden}, P.~M. 1979, \apj, 229, 942

\bibitem[{{Rosenberg} {et~al.}(2011){Rosenberg}, {Bern{\'e}}, {Boersma},
  {Allamandola}, \& {Tielens}}]{rosenberg_2011}
{Rosenberg}, M.~J.~F., {Bern{\'e}}, O., {Boersma}, C., {Allamandola}, L.~J., \&
  {Tielens}, A.~G.~G.~M. 2011, \aap, 532, A128

\bibitem[{{Salgado} {et~al.}(2012){Salgado}, {Bern{\'e}}, {Adams}, {Herter},
  {Gull}, {Schoenwald}, {Keller}, {De Buizer}, {Vacca}, {Becklin}, {Shuping},
  {Tielens}, \& {Zinnecker}}]{salgado_2012}
{Salgado}, F., {Bern{\'e}}, O., {Adams}, J.~D., {et~al.} 2012, \apjl, 749, L21

\bibitem[{{Sandstrom} {et~al.}(2012){Sandstrom}, {Bolatto}, {Bot}, {Draine},
  {Ingalls}, {Israel}, {Jackson}, {Leroy}, {Li}, {Rubio}, {Simon}, {Smith},
  {Stanimirovi{\'c}}, {Tielens}, \& {van Loon}}]{sandstrom_2012}
{Sandstrom}, K.~M., {Bolatto}, A.~D., {Bot}, C., {et~al.} 2012, \apj, 744, 20

\bibitem[{{Schaerer} \& {de Koter}(1997)}]{schaerer_1997}
{Schaerer}, D. \& {de Koter}, A. 1997, \aap, 322, 598

\bibitem[{{Sim{\'o}n-D{\'{\i}}az} {et~al.}(2011){Sim{\'o}n-D{\'{\i}}az},
  {Caballero}, \& {Lorenzo}}]{simon_diaz_2011}
{Sim{\'o}n-D{\'{\i}}az}, S., {Caballero}, J.~A., \& {Lorenzo}, J. 2011, \apj,
  742, 55

\bibitem[{{Smith} {et~al.}(2007){Smith}, {Armus}, {Dale}, {Roussel}, {Sheth},
  {Buckalew}, {Jarrett}, {Helou}, \& {Kennicutt}}]{smith_2007}
{Smith}, J.~D.~T., {Armus}, L., {Dale}, D.~A., {et~al.} 2007, \pasp, 119, 1133

\bibitem[{{Spitzer}(1978)}]{spitzer_1978}
{Spitzer}, Jr., L. 1978, \jrasc, 72, 349

\bibitem[{{Steinacker} {et~al.}(2010){Steinacker}, {Pagani}, {Bacmann}, \&
  {Guieu}}]{steinacker_2010}
{Steinacker}, J., {Pagani}, L., {Bacmann}, A., \& {Guieu}, S. 2010, \aap, 511,
  A9

\bibitem[{{Szczepanski} \& {Vala}(1993)}]{szczepanski_1993}
{Szczepanski}, J. \& {Vala}, M. 1993, \apj, 414, 646

\bibitem[{{Tenorio-Tagle}(1979)}]{tenorio_tagle_1979}
{Tenorio-Tagle}, G. 1979, \aap, 71, 59

\bibitem[{{Tielens}(1983)}]{tielens_1983}
{Tielens}, A.~G.~G.~M. 1983, \apj, 271, 702

\bibitem[{{Tielens}(2005)}]{tielens_2005}
{Tielens}, A.~G.~G.~M. 2005, {The Physics and Chemistry of the Interstellar
  Medium}

\bibitem[{{Tielens}(2008)}]{tielens_2008}
{Tielens}, A.~G.~G.~M. 2008, \araa, 46, 289

\bibitem[{{van Boekel} {et~al.}(2003){van Boekel}, {Waters}, {Dominik},
  {Bouwman}, {de Koter}, {Dullemond}, \& {Paresce}}]{van_boekel_2003}
{van Boekel}, R., {Waters}, L.~B.~F.~M., {Dominik}, C., {et~al.} 2003, \aap,
  400, L21

\bibitem[{{van Buren}(1986)}]{van_buren_1990}
{van Buren}, D. 1986, \apj, 306, 538

\bibitem[{{van Leeuwen}(2007)}]{van_leeuwen_2007}
{van Leeuwen}, F. 2007, \aap, 474, 653

\bibitem[{{van Loon} \& {Oliveira}(2003)}]{van_loon_2003}
{van Loon}, J.~T. \& {Oliveira}, J.~M. 2003, \aap, 405, L33

\bibitem[{{Veilleux} {et~al.}(2005){Veilleux}, {Cecil}, \&
  {Bland-Hawthorn}}]{veilleux_2005}
{Veilleux}, S., {Cecil}, G., \& {Bland-Hawthorn}, J. 2005, \araa, 43, 769

\bibitem[{{Veneziani} {et~al.}(2010){Veneziani}, {Ade}, {Bock}, {Boscaleri},
  {Crill}, {de Bernardis}, {De Gasperis}, {de Oliveira-Costa}, {De Troia}, {Di
  Stefano}, {Ganga}, {Jones}, {Kisner}, {Lange}, {MacTavish}, {Masi},
  {Mauskopf}, {Montroy}, {Natoli}, {Netterfield}, {Pascale}, {Piacentini},
  {Pietrobon}, {Polenta}, {Ricciardi}, {Romeo}, \& {Ruhl}}]{veneziani_2010}
{Veneziani}, M., {Ade}, P.~A.~R., {Bock}, J.~J., {et~al.} 2010, \apj, 713, 959

\bibitem[{{Warren} \& {Hesser}(1978)}]{warren_1978}
{Warren}, Jr., W.~H. \& {Hesser}, J.~E. 1978, \apjs, 36, 497

\bibitem[{{Weingartner} \& {Draine}(2001)}]{weingartner_2001}
{Weingartner}, J.~C. \& {Draine}, B.~T. 2001, \apj, 548, 296

\end{thebibliography}

\end{document}